%% file: manuscript_v2/science_template.tex
\documentclass[12pt]{article}
\usepackage{newtxtext,newtxmath}
\usepackage[utf8]{inputenc}

\usepackage{graphicx}

\usepackage[letterpaper,margin=1in]{geometry}

\renewenvironment{abstract}
	{\quotation}
	{\endquotation}

\date{}

\makeatletter
\renewcommand{\fnum@figure}{\textbf{Figure \thefigure}}
\renewcommand{\fnum@table}{\textbf{Table \thetable}}
\makeatother

\usepackage{manuscript_v2/scicite}

\usepackage{url}

\def\scititle{
    Two Protons, Two Positrons, and Four Electrons: Covalent Bond with van der Waals Characteristics
}
\title{\bfseries \boldmath \scititle}

\author{
	Jorge Charry$^{1,2,3}$,
	Alexandre Tkatchenko$^{1}$\and
	\small$^{1}$Department of Physics and Materials Science, University of Luxembourg, L-1511 Luxembourg City, Luxembourg.\and
	\small$^{2}$Luxembourg Researchers Hub asbl, 223 rue de Luxembourg, L-4222, Esch-sur-Alzette, Luxembourg.\and
   	\small$^{3}$HPC Platform, University of Luxembourg, L-4364 Esch-sur-Alzette, Luxembourg.\and
	\small$^\ast$Corresponding author. Email: jorge.charry@researchers-hub.lu, alexandre.tkatchenko@uni.lu \and
}

\begin{document} 

\maketitle

\begin{abstract} \bfseries \boldmath
Classifying interactions is key in the physical sciences, and bonding mechanisms in matter-antimatter systems remain particularly enigmatic. 
Here we focus on a paradigmatic example of positronium hydride (PsH) dimer composed of two protons, two positrons, and four electrons, whose bonding nature has been previously described as either ionic, covalent, or van der Waals-like. Accurate quantum Monte Carlo calculations show that the two positrons occupy a delocalized molecular orbital that envelopes the two hydrogen anions and responds as a collective dipole to an applied electric field.
This positronic bonding stems from quantum correlations that resemble a single covalent bond formed between negatively charged pseudo-nuclei, but with a bond strength commensurate with the traditional van der Waals interaction. 
Thus, polarization response serves as a diagnostic tool complementary to traditional charge density analysis that enables a robust description even for positron bonding. 
Our findings suggest that the ability to form delocalized proto-bonds is a more general property of quantum systems, and could be present in a broader class of particles, antiparticles, and quasi-particles interacting with matter. 
\end{abstract}

\section*{INTRODUCTION}
Understanding the interaction of positrons with matter and fields is important for a wide range of scales, from cosmological to subatomic. The range of outstanding puzzles involving positrons includes the mechanism for asymmetry between matter and antimatter in the early universe~\cite{Baryogenesis-RMP} to the computation of electron--positron field polarization in quantum electrodynamics~\cite{Tkatchenko-PRL2023}. At the atomic scale, the detection of gamma rays produced during the electron-positron pair annihilation process is the basis for useful technological applications, such as Positron Emission Tomography (PET) to map biological activity in tissues \cite{bass_RMP_95_021002_2023} and Positron Annihilation Lifetime Spectroscopy (PALS) to probe the (im)purity of materials \cite{gidley_ARMR_36_49_2006,singh_ASR_51_359_2016}.

More recently, positrons entered molecular sciences as well.
Low-energy positrons\cite{Anderson1933,schippers_JPBAMOP_52_171002_2019} can form stable states with electrons, atoms, and molecules prior to the pair annihilation process. 
Experimental detection via the vibrational Feshbach resonance (VFR) technique has confirmed positron binding in roughly 100 molecules, with observed binding energies reaching 300 meV \cite{danielson_PRA_111_042809_2025}. 
These molecular–positron complexes have attracted significant interest because their relatively long lifetimes (on the order of nanosecond) allow them to interact with nuclear motion in molecules, which typically unfolds on femtosecond to picosecond time scales\cite{alfredodupasquier____2010}. 
As a result, positrons can influence or alter molecular reaction pathways. Furthermore, because annihilation rates are intrinsically linked to the molecular environment, new positronium imaging techniques have emerged to exploit this sensitivity \cite{moskal_SA_7_eabh4394_2021}, extracting more biological and medically relevant information within living tissues that complements standard PET. 

Therefore, in order to discover and understand novel positronic chemistry, robust theoretical tools are needed to accurately model interacting matter--antimatter systems. A critical challenge is to properly account for electron–-positron quantum correlation effects~\cite{Hofierka2022, cassella_NC_15_5214_2024, upadhyay_JCTC_20_9879_2024}.
In recent decades, most theoretical studies have focused on the binding of positrons to stable molecules \cite{Bressanini1998, charlton____2001, Jean2003, chojnacki_ECWFiCaP__439_2003, alfredodupasquier____2010, Gribakin2010, kita2012, schippers_JPBAMOP_52_171002_2019, Hofierka2022, cassella_NC_15_5214_2024}.  
It was also found that positrons can stabilize the interaction between two or more otherwise repelling atoms or ions, forming a new type of bond rather than simply attaching to a molecule.
Starting from the case of one positron between two hydrogen anions H$^-$e$^+$H $^-$ \cite{charry_AC-IE_57_8859_2018}, to other single positron bonds in halides \cite{moncada_CS_11_44_2020,tachibana_PRL_131_143201_2023}, beryllium \cite{upadhyay_JCTC_20_9879_2024, porras-roldan_CS_16_22322_2025} and molecular anions dimers \cite{cassidy_JCP_160_084304_2024}, the case of two positrons in two hydrides\cite{bressanini_JCP_155_054306_2021}, halides \cite{archila-pena_CEJ_30_202402618_2024}, three hydride centers \cite{charry_CS_13_13795_2022}, and even three-positron two hydride system\cite{bressanini_JCP_156_154302_2022}. 

One of the simplest paradigmatic cases of the proton-electron-positron-containing molecule is positronium hydride (PsH). Its ground state consists of hydrogen anion \textit{dressed} with a positron in its outer shell. Because PsH is a neutral moiety, it has been proposed that the interaction between two PsH molecules is purely non-covalent \cite{Mella2001, Yan2002}. However, Bressanini~\cite{bressanini_JCP_155_054306_2021} used quantum Monte Carlo calculations to demonstrate that at an H-H distance of 6.0 Bohr, the two positrons in the PsH dimer bind two otherwise mutually repulsive hydrogen anions with an interaction energy of 6.5 kcal/mol. This is similar to the single-positron case, which, however, binds two hydrogen anions by 14.8 kcal/mol~\cite{charry_AC-IE_57_8859_2018}. 
It is thus clear that the nature of unique bonding in the PsH dimer is unsolved \cite{bressanini_JCP_155_054306_2021, goli_PCCP_25_29531_2023, goli_pccp_28_11154_2026}, shifting from covalent as in H$_2$ or Li$_2$, three-center ionic-like as in $q^-2q^+q^-$ model, back to a ``super van der Waals'' description driven by strong quantum correlations, all with unique features compared to regular electronic bonds.

\begin{figure*}
    \centering
    \includegraphics[width=1\linewidth]{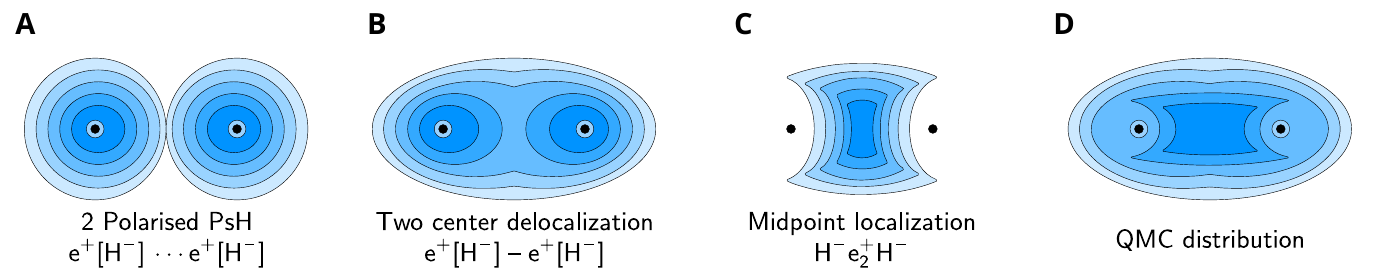}
    \caption{ \textbf{Schematic comparison for possible bond mechanism interpretations in (PsH)$_2$.} \\
    Simplified contour distribution for the positronic density distribution in (PsH)$_2$. \textbf{(A)} Two isolated PsH polarizing each other and interacting via van-der-Waals mechanism. \textbf{(B)} Expected distribution from electronic covalent bond analogues. \textbf{(C)} Localized three-center ionic-like distribution. \textbf{(D)}  
    QMC positronic density distribution at the equilibrium geometry. 
    } 
    \label{fig:psh2_schematic}
\end{figure*}

A simplified scheme for these different proposed mechanisms is shown in Fig. \ref{fig:psh2_schematic}.
Ultimately, these overlapping yet distinct features indicate a bonding mechanism that defies standard classification, leaving the true character of the bond an open puzzle. We propose that this situation arises because the traditional chemical bond analysis tools based on particle densities and bond strengths are insufficient to explain interactions in matter--antimatter systems.
Instead, we offer a solution by performing converged and quantitative quantum Monte Carlo calculations and analyzing the polarization response properties of the PsH dimer. QMC is particularly suited for this task as it allows for the integration of explicitly correlated wave functions \cite{Bressanini1998, Kita2010, charrymartinez_JCTC_18_2267_2022, marsusi_PRA_106_62822_2022, simula_PRL_129_166403_2022, cassella_NC_15_5214_2024, upadhyay_JCTC_20_9879_2024, barborini_JCP_164_062501_2026}, which are essential for accurately capturing the complex correlation between positrons and molecular systems. 
By computing (PsH)$_2$ polarizability as a function of the internuclear distance and applying the polarizability decomposition analysis recently introduced in reference \citenum{charry_PCCP_27_23044_2025}, we aim to provide a comprehensive analysis on the characterization of bonding in (PsH)$_2$. We start by deconstructing the traditional bonding mechanisms.

\section*{RESULTS}
\subsection*{(PsH)$_2$: not a traditional chemical bond}

The interaction in (PsH)$_2$, and other positron bonds, has been previously analyzed by the following approaches: (i) comparing the properties of its potential energy curve against regular electronic systems; (ii)  analyzing the topology of the electronic and positronic density; (iii) performing energy decompositions commonly expressed in terms of electrostatic, polarization, exchange, and correlation components.

Although H$_2$ is the standard model for chemical bonding, positron-bonded systems share features closer to alkali metals, specifically in their internuclear distances and bond strengths, even though their underlying physics is clearly different \cite{charry_AC-IE_57_8859_2018, moncada_CS_11_44_2020, charry_CS_13_13795_2022, archila-pena_CEJ_30_202402618_2024, bressanini_JCP_155_054306_2021, goli_PCCP_25_29531_2023}.

\begin{figure}[tpb!]
\centering
\includegraphics[width=0.6\textwidth]{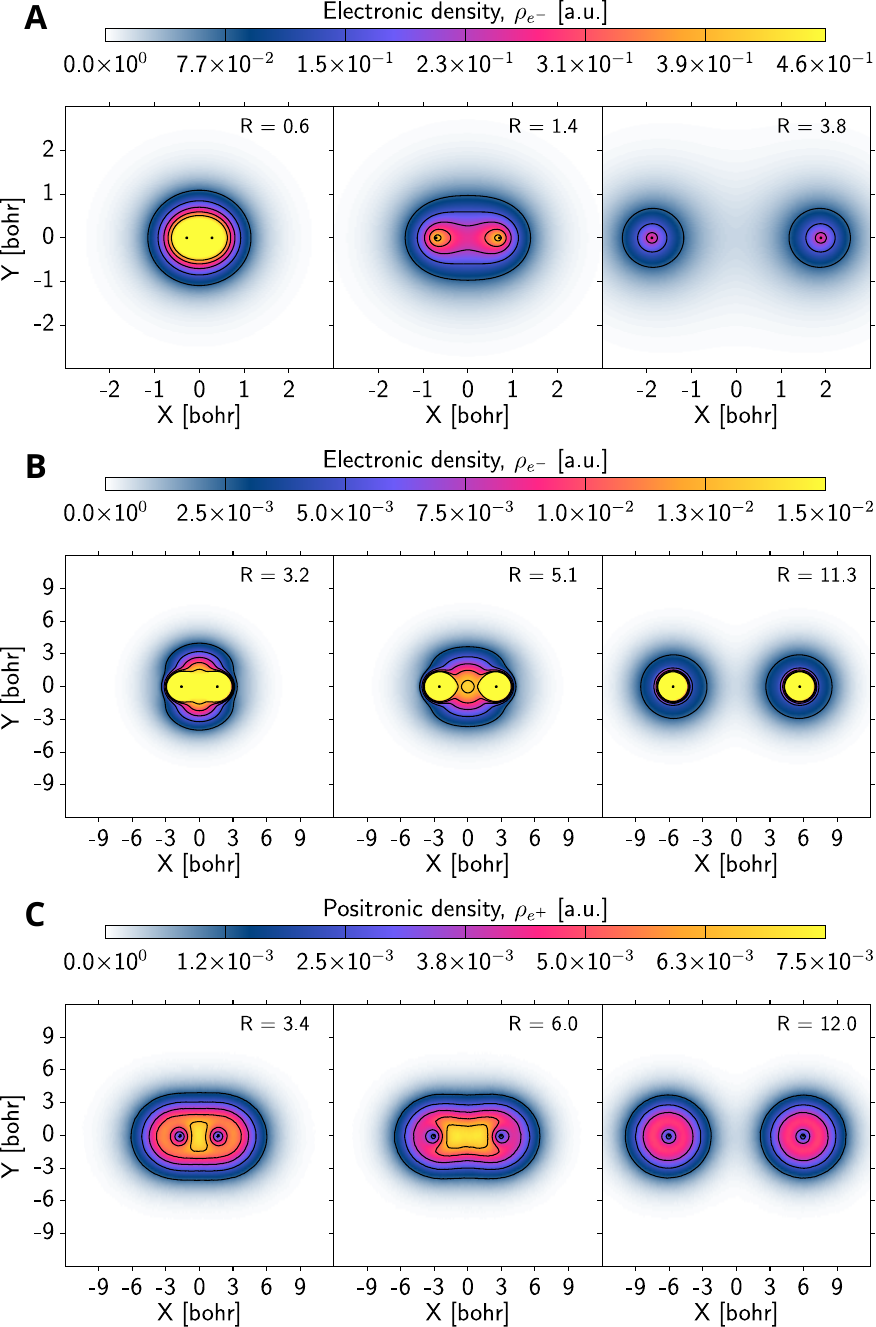}
\caption[PsH2 density]{ \textbf{Comparison of bonding densities in diatomic molecules.} \\ CCSD natural orbital electronic density for H$_2$ (\textbf{A}) and Li$_2$ (\textbf{B}). DMC Positronic density for (PsH)$_2$ (\textbf{C}). Each panel left, mid, and right corresponds to electronic or positronic densities computed at three representative internuclear separations: short range (near the dissociation channel Ps$_2$ + H$_2$ for (PsH)$_2$), equilibrium distance, and dissociation limit.}
\label{fig:psh2_dens}
\end{figure}

\begin{figure}[tpb!]
    \centering
    \includegraphics[width=0.45\linewidth]{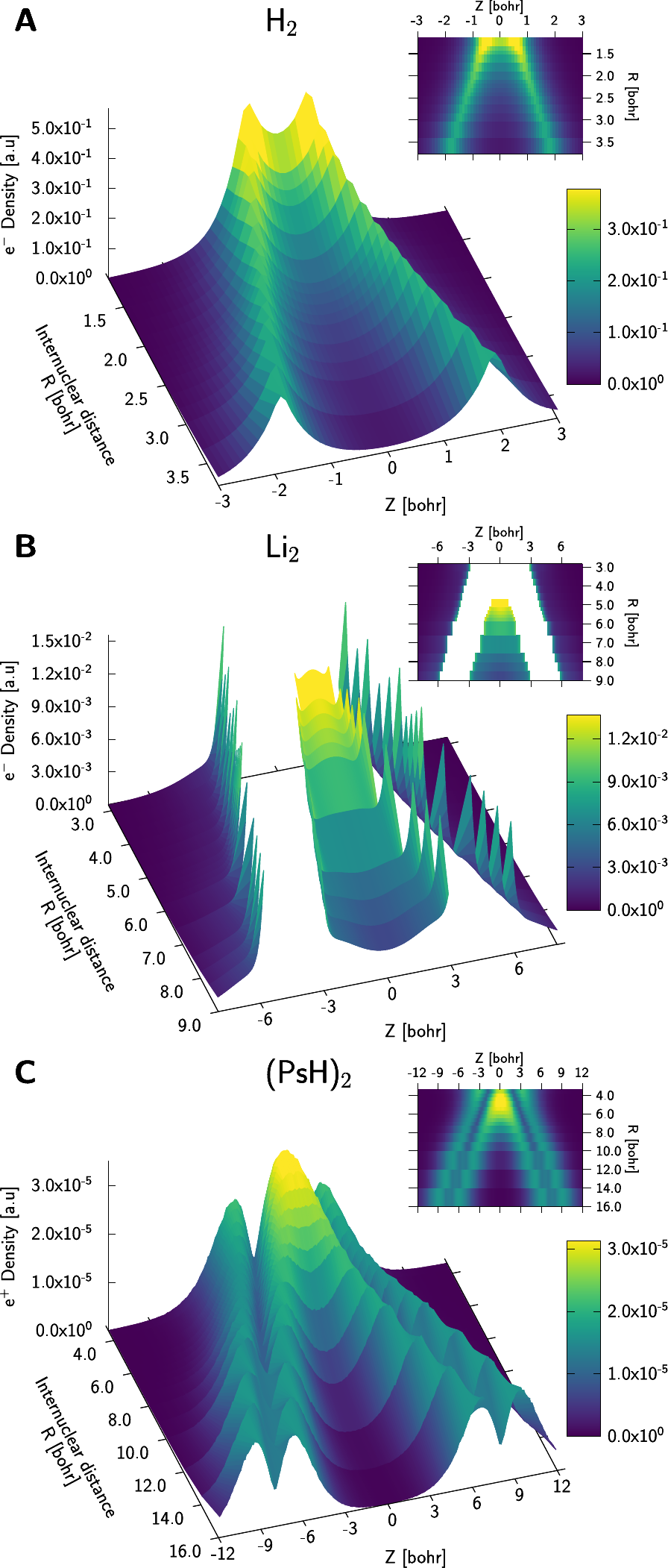}
    \caption{ \textbf{One-dimensional bonding density evolution.} \\ Interpolated surfaces of density slices on the $xy$ plane parallel to the bond axis ($z$-axis), showing their evolution with respect to the internuclear distance. 
CCSD natural orbital electronic density for H$_2$ (\textbf{A}) and Li$_2$ (\textbf{B}). DMC Positronic density for (PsH)$_2$ (\textbf{C}).}
\label{fig:psh2_1d_dens_evolution}
\end{figure}

Given the above similarities, the bonding densities for electronic and positronic systems have also been compared and analyzed in detail \cite{bressanini_JCP_155_054306_2021, goli_PCCP_25_29531_2023}, shown here in Fig. \ref{fig:psh2_dens} as 2D density plots (And its Laplacian in SM), and in Fig. \ref{fig:psh2_1d_dens_evolution} as evolution of 1D density slices as a function of the internuclear distance.
As reported before, by comparing against the densities of two non-interacting PsH, the electronic density in (PsH)$_2$ remains largely unaffected upon the addition of the positron \cite{goli_PCCP_25_29531_2023}.
On the other hand, the positron pair seems to serve as the primary mediator of the bond since its major charge density redistribution from one PsH to the dimer. 
Although the positronic distribution surrounding the two H$^-$ anions mimics the appearance of a covalent bond, this is insufficient to prove a traditional covalent character due to several unique topological features compared with those of H$_2$ and Li$_2$.
Specifically, the positronic distribution exhibits a ``hole'' at the nuclear positions instead of the standard cusp, a result of the electrostatic repulsion between the positrons and the nuclei. 
Moreover, the density is mostly localized in the middle of the two ions, reaching a maximum along the internuclear axis and also along the perpendicular plane. 
Despite the statistical noise, the corresponding positronic Laplacian density profile shows a clear trend toward accumulation regions at the midpoint between the two $\text{H}^-$ units, with depletion near the nuclei due to the positron repulsion as reported before \cite{goli_PCCP_25_29531_2023} (See also in Fig S1).
As seen in Fig. \ref{fig:psh2_1d_dens_evolution}, at large distances, the positronic density exhibits two distinct maxima per PsH, localized around each hydride center, with their corresponding minima at the nucleus position due to the positron-nuclei repulsion. 
As the fragments approach the equilibrium distance, the two internuclear maxima fuse into a single prominent maximum at the bond midpoint.
This topology mirrors the emergence of a "pseudoatom" or Non-Nuclear Attractors (NNAs) bridging the two centers, characteristic of loosely bound, highly polarizable quantum particles.
A feature present in molecular electrides \cite{postils_CC_51_4865_2015, sitkiewicz_JPCA_125_4819_2021} and in Li$_2$ at a much lower density scale, but absent in $H_2$.

Due to the high electrostatic attractive interaction, the concentration of positive charge ($2q^+$) symmetrically localized between the two negatively charged hydrogen anion centers ($q^-$), could suggest a three-center ionic-like character of the bond as $q^-2q^+q^-$. 
Not to be confused with the ionic bond $q^-q^+$ in standard electronic molecular systems. 
However, energy decomposition analysis \cite{goli_PCCP_25_29531_2023,goli_pccp_28_11154_2026} indicated that such attraction is insufficient, instead the stabilization is mainly due to electron-positron and positron-positron correlation effects.
Since correlation driven interactions between neutrally charged atoms are commonly associated with dispersion interactions, Goli et al. \cite{goli_pccp_28_11154_2026} compared the binding energy of (PsH)$_2$ with the dispersion energy estimated from previously computed van der Waals C$_6$ and C$_8$ coefficients of isolated PsH\cite{Yan2002} as $E_{\text{disp}}^{\text{(PsH-PsH)}} \approx -\frac{C_6}{R^6} - \frac{C_8}{R^8}$. 
Remarkably, for PsH the ratio of dispersion coefficients is relatively high, $C_8/C_6 \approx 90$, which is comparable to heavy alkali ( Na-Na: 74 -- Cs-Cs: 149 ) or alkaline earth metals ( Sr-Sr: 68 -- Ba-Ba: 91), in contrast to lower ratios within first and second period ( H-H: 19, Li-Li : 60 ) or noble gases ( He-He: 10 -- Xe-Xe: 42). 
As a consequence, at the equilibrium distance, $E_{\text{disp}}^{\text{(PsH-PsH)}} = -12.9$ kcal/mol roughly captures the order of magnitude of the DMC binding energy of -6.5 kcal/mol, but with a 70$\%$ contribution from C$_8$ and 30$\%$ from C$_6$ (See SM).
Therefore, given the relatively high binding energy compared to regular electronic vdW interactions (1-4 kcal/mol), the prominent role of positron-positron correlation, and the large dispersion coefficients, it was suggested to denote the bond as a ``super van der Waals'' bond \cite{goli_pccp_28_11154_2026}.

\subsection*{(PsH)$_2$: not a van-der-Waals bond}
\begin{figure}[tpb!]
\centering
\includegraphics[width=1.0\textwidth]{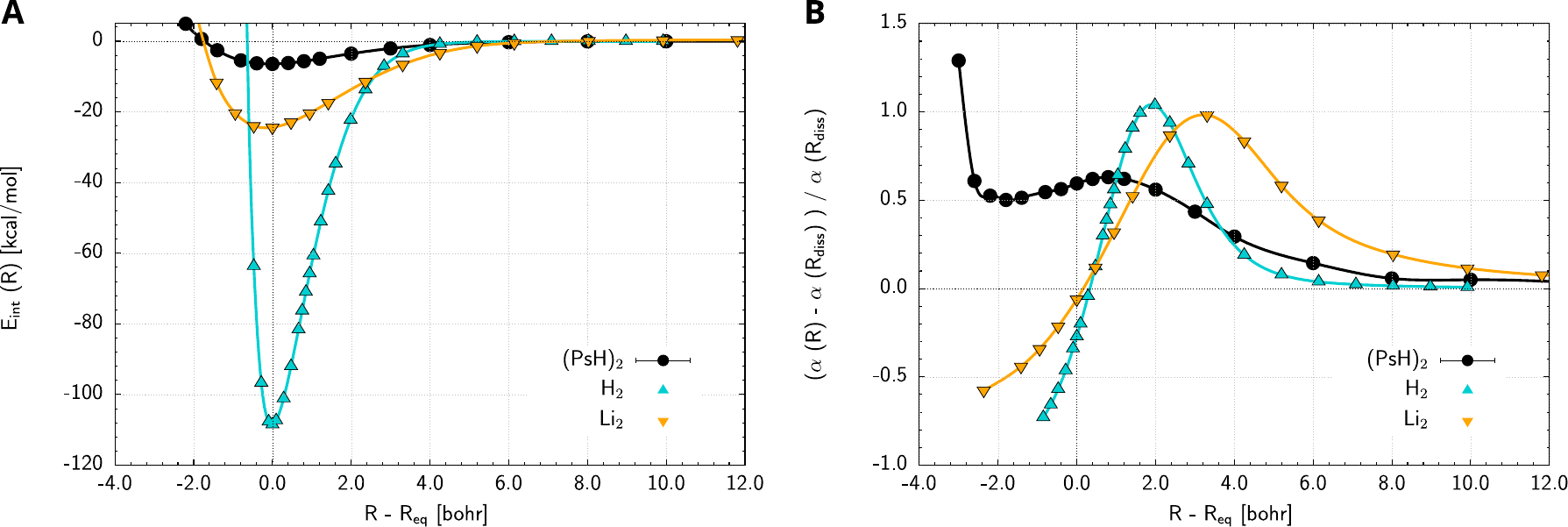}
\caption[PsH2 PES]{ \textbf{Interaction energies and polarizabilities curves for diatomic molecules} \\
\textbf{(A)} Interaction energies (E$_{\text{AB}}$(R) 
- E$_{\text{A}}$ - E$_{\text{B}}$) and \textbf{(B)}
relative polarizability (parallel component with respect to the bond axis) changes as a function of the internuclear separation shifted to the equilibrium for PsH$_2$, H$_2$, and Li$_2$.}
\label{fig:psh2_pes}
\end{figure}

The first argument against a van-der-Waals bond is that this interpretation contradicts the large reorganization of the positronic density from PsH to (PsH)$_2$ (Fig. 2). This opposes the picture of two isolated atoms interacting through instantaneous dipoles required for a vdW dispersion interaction. In addition, the DMC energy curve decays slower than the vdW $R^{-6}$ potential.

\begin{figure}[htbp!]
    \centering
    \includegraphics[width=0.6\linewidth]{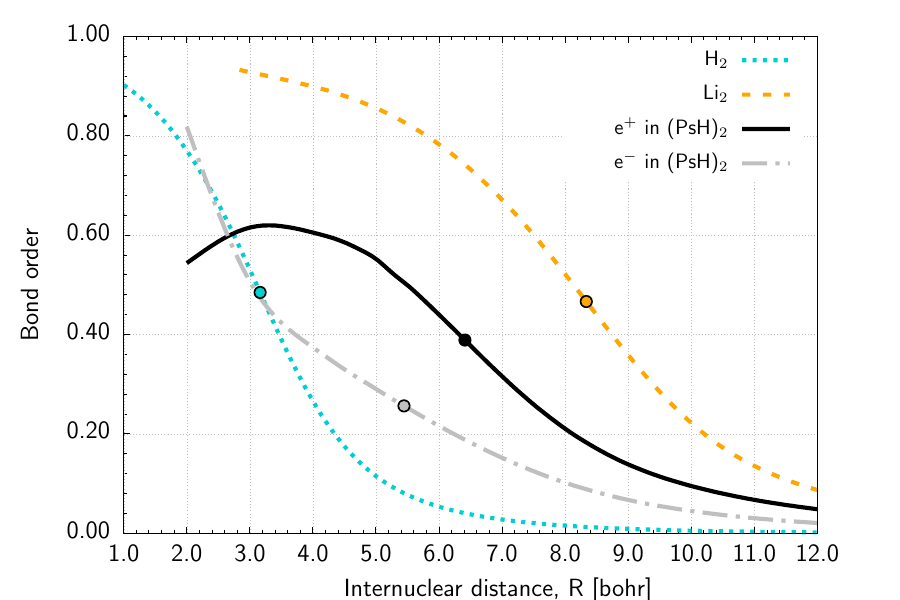}
    \caption{
    Delocalization Index (DI) bond order for H$_2$, Li$_2$, and positronic and electronic DI in (PsH)$_2$. The data for H$_2$ and Li$_2$ were taken from the open Zenodo repository of reference Charry PCCP 2025 \cite{charry_PCCP_27_23044_2025}. Circles are used to highlight the identification of an inflexion point.}
    \label{fig:DI_bond_order}
\end{figure}

To quantify particle sharing, we computed in Fig. \ref{fig:DI_bond_order} the delocalization indices (DI)  \cite{angyan_CPL_299_1_1999, Outeiral2018, Silva2020} using a multicomponent compacted active-space Full Configuration Interaction (FCI) approach \cite{bollhofer_CPC_177_951_2007, Flores-Moreno2014, charry_AC-IE_57_8859_2018} validated against our QMC potential energy surface (see SM). 
Around the equilibrium distance, the positronic DI is significantly larger than its electronic counterpart and exhibits a characteristic sigmoid decay upon dissociation, a characteristic of shared, covalent interactions analogous to $\text{Li}_2$ and $\text{H}_2$ \cite{Garcia-Revilla2011, charry_PCCP_27_23044_2025}. 
Conversely, the electronic DI is roughly half the magnitude and decays exponentially, indicating a non-bonded character \cite{Garcia-Revilla2011} and confirming that the positronic cloud acts as the primary collective glue in the dimer.
A broader study across a wider array of multi-positronic complexes remains an open avenue to fully map these unique DI topologies.

As additional evidence, we present in Figure \ref{fig:psh2_pes} the evolution of the energy and relative polarizability along the internuclear distance of (PsH)$_2$.
We also include data from H$_2$ and Li$_2$, systems previously used for comparison with positronic bonds \cite{hait_AC_62_e202312078_2023, charry_PCCP_27_23044_2025}. 
As seen from there, the strength of the (PsH)$_2$ bond is close to the Li$_2$ case, which is considered as a weak covalent bond \cite{sterling_JACS_146_9532_2024}.
The polarizability values were derived from both the first-order derivative of the dipole moment and the second-order derivatives of the energy (refer to the SM for details on our methodology).
Our results show that the polarizability (130.5(2) bohr$^3$) — computed along the axis connecting the hydrogen nuclei — is considerably higher than the sum of two non-interacting PsH monomers (84.6 bohr$^3$). 
This enhancement indicates a substantial redistribution of the wave function in comparison to that of the isolated monomers. 
This behavior mirrors the polarizability trends observed in covalently bonded diatomic molecules at their equilibrium distances \cite{Rychlewski1980, Gopakumar2014, Miliordos2018, hait_AC_62_e202312078_2023, charry_PCCP_27_23044_2025}.

Clearly, at shorter distances, when the system factorizes in the sum of Ps$_2$ and H$_2$ \cite{bressanini_JCP_155_054306_2021}, the polarizability abruptly increases, as expected, but this regime is not the focus of this study.

\begin{figure}[tpb!]
\centering
\includegraphics[width=1.0\textwidth]{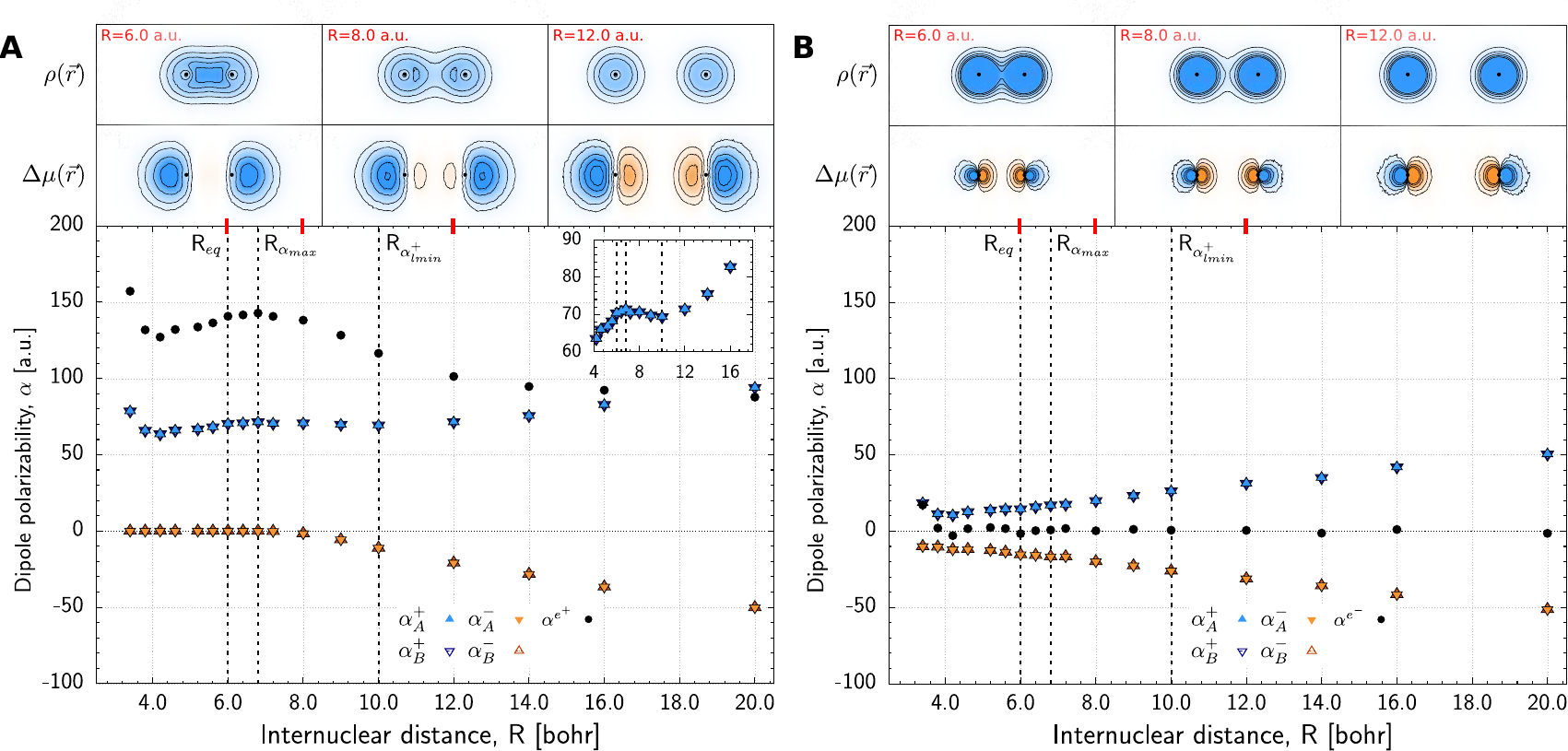}
\caption[PsH2 decomp]{
\textbf{Dipole response under an applied electric field on PsH$_2$.} \\
Spatial decomposition curves of positronic ({\bf A}) and electronic ({\bf B}) dipole polarizabilities according to eq. \ref{eq:mu_pol}, as illustrated in the top panels for the 2D local density $\rho$ and dipole variation $\Delta\mu$ plots at representative internuclear separations. 
Orange and blue regions correspond to a decrease and an increase of positronic or electronic local quantities, respectively.
Dotted vertical lines correspond to equilibrium distance $R_{eq}$, maximum polarizability peak $R_{\alpha_{max}}$, and local minima of $\alpha^+_{lmin}$ polarizability component $R_{\alpha^+_{lmin}}$. The (\textbf{A}) inset shows a zoomed-in region near the critical points on $\alpha^+$ curves. 
} 
\label{fig:psh2_decom}
\end{figure}

To shed further light on the characteristics of the polarizability, we decompose the dipole moment operator into electronic $\boldsymbol{\mu}^{e} = -\sum^{N_{e}}_{i=1} \textbf{r}_{i}$ and $\boldsymbol{\mu}^{p} = \sum^{N_{p}}_{i=1} \textbf{r}_{i}$ positronic parts, allowing to compute separate polarizability contributions of the two leptonic particles
\begin{equation}
\alpha_{ab} 
= \left . \frac{\partial \mu_a^{e}(F_b) }{\partial F_b}\right |_{F_b =0} 
 + \left . \frac{\partial \mu_a^{p}(F_b) }{\partial F_b}\right |_{F_b =0} 
= \alpha_{ab}^{e} + \alpha_{ab}^{p} 
\qquad a,b \in [x,y,z].
\end{equation}
The black curves in Figure \ref{fig:psh2_decom}a-b show the contribution of positronic and electronic polarizability, respectively. 
At all internuclear separations of (PsH)$_2$, the largest contribution comes from the delocalized positrons, while the electronic response is almost negligible. 
These results suggest that from the point of view of its response properties, a PsH system can be interpreted as a positronic cloud that fully screens the electronic response of the H$^-$ anion it surrounds. 
In fact, an isolated H$^-$ anion exhibits a large polarizability of 203 a.u. \cite{Mella2001}, which stems from loosely bound electrons that can easily be distorted. 
This observation contradicts the arguments of Yan \textit{et al.}\cite{Yan2002}, who proposed a simple estimate of the polarizabilities of PsH (42.3 a.u.) as a weakly bound state between Ps and an H atom (40.5 =  36 + 4.5 a.u.), and confirms later studies based on the wave functions distributions, which conclude that the structure of PsH is that of an H anion slightly perturbed by the positron\cite{Bressanini2003}.
The screened H$^-$ anion behavior seen in PsH is clearly also visible in (PsH)$_2$ at all distances.

In order to better understand the origin of the polarizability changes, we employed the local decomposition proposed by us in Ref.~\citenum{charry_PCCP_27_23044_2025}, which consists in computing the local polarizabilities of the two fragments $A$ and $B$ that participate in the bond, separating them into an accumulation(+) and a depletion(-) region, such that
\begin{equation}
\alpha_{zz} =\alpha^{+}_{A} + \alpha^{-}_{A} + \alpha^{+}_{B} +\alpha^{-}_{B},
\label{eq:mu_pol}
\end{equation}
where 
\begin{equation}
\left \{
\begin{array}{c}
\alpha^+_{\Omega}  = \int_{V_\Omega}\left . \max \left \{ 0, r_z \frac{\partial \rho(\vec{r},F_z) }{
\partial F_z} \right|_{F_z=0} \right \} d\vec{r} \\
\alpha^-_{\Omega}  = \int_{V_\Omega}\left . \min \left \{ 0, r_z \frac{\partial \rho(\vec{r},F_z) }{\partial F_z} \right|_{F_z=0} \right \} d\vec{r} \\
\end{array} \right . 
\label{eq:mu_pol_terms}
\end{equation}
with $\Omega \in [A,B]$, and assuming that the $z$ axis corresponds with the direction of the bond.
It is worth noting that while the polarizability decomposition employed here is fundamentally origin-dependent, a feature discussed in detail in our previous work \cite{charry_PCCP_27_23044_2025}, this does not introduce any ambiguity for (PsH)$_2$ due to its spatial symmetry. 
In essence, the bond center is uniquely defined as the exact geometric midpoint, which simultaneously coincides with the center of mass, the center of charge, and the topological bond critical point.
On the other hand, this introduces a linear dependency with the internuclear systems, which allows for interpreting the different bonded, bond-breaking, and dissociation regions under the context of the midpoint origin-dependency. 

Figure \ref{fig:psh2_decom} shows the above decomposition for the local variations of the positronic (a) and electronic (b) polarizabilities along the (PsH)$_2$ bond.
In the panels above, we also include a positronic density contour map and a local dipole variation at representative distances for an easier interpretation. 

An analysis of the individual accumulation and depletion components around fragments A and B reveals a distinct pattern. 
The local electronic polarizability components—denoted as $\alpha^+_A, \alpha^+_B$ for accumulation and $\alpha^-_A, \alpha^-_B$ for depletion—scale linearly with the internuclear distance (Figure \ref{fig:psh2_decom}b),
due to the formation of two dipoles in opposite directions that cancel out, leading to an almost zero total electronic dipole polarizability.
This behavior mirrors a "pseudo" non-covalent interaction, suggesting that the electrons remain localized around their respective nuclei with no significant exchange between the two anions \cite{charry_PCCP_27_23044_2025}.
Therefore, the unperturbed electronic response to an external field, in combination with the electronic static nature upon the addition of positrons reveals that electrons are not the bond mediators. 

In contrast, the positronic polarizability components exhibit a behavior characteristic of traditional covalent bonds \cite{charry_PCCP_27_23044_2025} (Figure \ref{fig:psh2_decom}a). 
At the equilibrium distance of 6.0 Bohr, an external electric field induces positronic density accumulation on the side of one nucleus facing away from the other, with a specular effect occurring at the opposite anion. 
Together, these local shifts establish a molecular dipole between hydrogen anions that evolves as a function of the internuclear distance.
As the fragments separate, this molecular dipole increases until the polarizability reaches its maximum ($\alpha_{max}$) at 7.2 Bohr. 
Beyond this peak, the dipole in the extra-bond regions ($\alpha_B^+, \alpha_A^+$) begins to diminish. 
Similarly, density reorganization in the intra-bond region generates an opposing dipole ($\alpha_B^-, \alpha_A^-$) that decreases linearly. 
A final transition occurs at 10 Bohr, marked by local minima in $\alpha_B^+$ and $\alpha_A^+$. 
This threshold identifies the onset of the dissociation region, where $\alpha_B^+$ and $\alpha_A^+$ resume a linear increase, similar to the trend seen in the intra-bond region. 
In this regime, the electric field induces two collinear atomic dipoles—one at each hydrogen nucleus—signaling the dissociation limit where the dimer's polarizability converges to the sum of two non-interacting PsH fragments.

\subsection*{ (PsH)$_2$: proto-bond with van-der-Waals strength}

The leading-order positronic polarizability curve of the PsH dimer closely mirrors that of a weak electronic covalent bond, as observed in typical diatomic systems \cite{hait_AC_62_e202312078_2023, charry_PCCP_27_23044_2025}.
In contrast, the much smaller electronic response remains largely static, effectively screened by the positronic cloud. 
From this perspective, this system is best described as a positronic distribution surrounding relatively unperturbed H$^-$ anions.
This similarity in the local dipole response is particularly remarkable given the unique positronic density distribution of (PsH)$_2$, which is highly localized at the bond’s midpoint, a feature that might superficially suggest an ionic-like bond $q^-2q^+q^-$ character, but the underlying physics tells a different story.
Such density distribution suggests an intriguing parallel to non-nuclear attractors in molecular electrides \cite{postils_CC_51_4865_2015, sitkiewicz_JPCA_125_4819_2021}, this connection could inspire the design and exploration of more positron-bonded compounds.

The covalent-like behavior arises because the positronic pair mediating the bond occupies a shared single ``molecular orbital'' that also encapsulates the two negatively charged hydrogen atoms, responding as a single molecular dipole to an external electric field. 
In this framework, the two H$^-$ anions act as ``pseudo-nuclei'' with diffused electronic wave functions that remain essentially separated from one another.
This picture illustrates the unique features of the bond and contradicts the long-held and still discussed assumption that the PsH dimer bonds purely through enhanced van der Waals interactions, despite the relatively low binding energy of 6.5 kcal/mol compared to standard covalent bonds. 
Moreover, our results establish polarizability as a vital complement to density analysis, offering a comprehensive framework for characterizing chemical bonds.

\section*{DISCUSSION}
\label{sec:conclusions}

In conclusion, we have systematically evaluated the nature of the bonding interaction in the $\text{(PsH)}_2$ by contextualizing our new findings with existing perspectives from the literature. 
A remarkably consistent picture that (PsH)$_2$ is not a traditional chemical bond emerges by analyzing a diverse set of physical and chemical descriptors, spanning energetics, energy decomposition, electronic and positronic density topologies, their Laplacian and evolution along the bond axis. 

The electron-pair and positron-pair sharing through delocalization indices coupled with our novel analysis of electronic and positronic polarizabilities response firmly demonstrate that the $\text{(PsH)}_2$ is not a van-der-Waals bond, but rather is characterized by a delocalized (proto)-bond with distinct covalent signatures compared to traditional electronic covalent interaction, defying traditional classification. 
Yet, the low binding energy of 6.5 kcal/mol is closer in magnitude to the typical scale of 1-4 kcal/mol for van der Waals interaction, estimated from its dispersion coefficients, given the large polarizability of each PsH monomer. 
Considering that, we finally conclude that (PsH)$_2$ is a proto-bond with van-der-Waals strength.

Overall, the tools and analysis presented in this work provide a comprehensive understanding of intramolecular positronic interactions, which are gradually reshaping and expanding the community's perspective on the field of positronic chemistry. 
Not only resolves the historical ambiguity surrounding this exotic system but also establishes a rigorous framework for diagnosing non-classical bonding in quantum matter.

These findings broadens the scope of bonding topologies of quantum systems, paving the way for the exploration of unconventional bond-like chemistry in a broader class of particles, antiparticles, and quasi-particles interacting with matter, for example, muon $\mu^-$ in muon-catalyzed fusion \cite{toyama_SA_12_eaed3321_2026}, antimuon $\mu^+$ in vibrational bonding for Br$\mu^+$Br like compounds \cite{fleming_AC-IE_53_13706_2014}, or electron--hole pairs in materials.

\section*{METHODS}
The polarizability and electronic density for $\text{H}_2$ and $\text{Li}_2$ were taken from Ref. \citenum{charry_PCCP_27_23044_2025}, having been originally computed with Psi4 code \cite{smith_JCP_152_184108_2020} at the $\text{CCSD(T)}$ and $\text{CCSD}$ levels, respectively, using the $\text{aug-cc-pVTZ}$ basis set. 
The CC density was generated from the natural orbitals of the CCSD ground-state one-electron reduced density matrix (1RDM) using the Molden interface of the Psi4 code.
The multicomponent Full Configuration Interaction (FCI/aug-cc-pVTZ) was computed from a reduced active space selected from the natural orbitals of a preceding CISD calculation using the FCI implementation of the openLOWDIN code \cite{charry_AC-IE_57_8859_2018, Flores-Moreno2014} and Jadamilu library for the large sparse matrix diagonalization \cite{bollhofer_CPC_177_951_2007}. 

The detailed description of the quantum Monte Carlo (QMC) methods and wave function applied for this investigation is reported in the Supplementary Material (SM).
A single Slater determinant was employed for the electronic wave functions, built as a linear combination of atomic-centered Gaussian-type orbitals, and the positronic part of the wave function is always built through the EPO ansatz generalized in ref. \citenum{charrymartinez_JCTC_18_2267_2022}.
In order to properly describe all correlations in these systems, we have also introduced a novel Jastrow described in the SM.
For all the calculations, we apply a modified version of the QMeCha quantum Monte Carlo package\cite{barborini_JCP_164_062501_2026}.

The H atomic basis is built with (3s1p)/[1s1p]  contracted Gaussian orbitals. 
For simplicity, the same set of orbitals is used for both the positronic and fermionic wave determinantal parts.
The dynamical Jastrow basis is built from (3s2p) uncontracted and not normalized GTOs centered on each atom, and (3s2p) uncontracted electron-positron GTOs.
All the parameters of the wave function are optimized simultaneously, including the contracted exponents and coefficients, the molecular orbital coefficients, the cusps parameters, and the linear and orbital parameters of the dynamical Jastrow factor. 
Regarding the DMC calculations, each optimized VMC wave function was used as the guiding function, and the estimation of the DMC energy was performed with 6400 walkers divided into 5000 blocks, each 1000 steps long, for a time step of 0.001 hartree$^{-1}$. 

The QMC density plots were obtained by counting the number of particles of each weighted configuration in a three-dimensional grid of 101,101,401 points with a length step of 0.02, 0.02, 0.01 bohr. 
For all the electronic properties that do not commute with the Hamiltonian, the second order estimator (SOE) was used, for which $\langle\hat{A}\rangle_{\text{SOE}} \approx 2\langle\hat{A}\rangle_{\text {DMC}}-\langle\hat{A}\rangle_{\text {VMC}}$.
All response properties are computed through a polynomial fit using fields of variable strength in the interval [0.001, 0.007] (see SM for further details).
\clearpage
\newpage


\clearpage 

%
\bibliography{manuscript_v2/references} 
\bibliographystyle{manuscript_v2/sciencemag.bst}

%
%
%
%
%
%

\section*{Acknowledgments}
We thank Dr. Matteo Barborini for his valuable insights, mentorship, QMeCha code development, Jastrow factor implementation, supervision, and writing contribution to the initial drafting of this manuscript.
We also thank Dr. Neil Drummond and Dr. Michele Casula for fruitful discussions. 
\paragraph*{Funding:}
JC acknowledges financial support from the Luxembourg National Research Fund, FNR (AFR PhD/19/MS, 13590856).
The calculations presented in this paper were carried out using the HPC facilities of the University of Luxembourg~\cite{VBCG_HPCS14} {\small (see \url{http://hpc.uni.lu})} and those of the Luxembourg national supercomputer MeluXina.
The authors gratefully acknowledge both ULHPC and LuxProvide teams for their expert support.
\paragraph*{Author contributions:}
JC: Conceptualization, Data curation, Formal Analysis, Investigation, Methodology, Software, Visualization, Writing – original draft, Writing – review \& editing.
AT: Conceptualization, Funding acquisition, Resources, Supervision, Writing – review \& editing.
\paragraph*{Competing interests:}
There are no competing interests to declare.
\paragraph*{Data and materials availability:}
Data for this study, including raw results, wave function and density outputs, as well as postprocessing scripts, are available in the Zenodo archive \url{https://doi.org/10.5281/zenodo.20184717}.

\subsection*{Supplementary materials}
Supplementary Text\\
Figs. S1 to S3\\
Tables S1 to S4\\
References 47 to 75


\newpage


\renewcommand{\thefigure}{S\arabic{figure}}
\renewcommand{\thetable}{S\arabic{table}}
\renewcommand{\theequation}{S\arabic{equation}}
\renewcommand{\thepage}{S\arabic{page}}
\setcounter{figure}{0}
\setcounter{table}{0}
\setcounter{equation}{0}
\setcounter{page}{1} 


\begin{center}
\section*{Supplementary Materials for\\ \scititle}

Jorge Charry$^{1,2,3}$,
Alexandre Tkatchenko$^{2}$\\
\small$^\ast$Corresponding author. Email: jorge.charry@researchers-hub.lu, alexandre.tkatchenko@uni.lu \\
\end{center}

\subsubsection*{This PDF file includes:}
Supplementary Text\\
Figures S1 to S13\\
Tables S1 to S3\\
References 47 to 75


\newpage


\input{supp_info_v2/0_laplacian}
\input{supp_info_v2/1_dispersion}
\input{supp_info_v2/2_quantummontecarlo}
\input{supp_info_v2/3_hamiltonian}
\input{supp_info_v2/4_wavefunction}
\input{supp_info_v2/5_Jastrow}
\input{supp_info_v2/6_polarizabilities}
\input{supp_info_v2/7_annihilation}
\input{supp_info_v2/8_general-results}
\input{supp_info_v2/9_PES}

\end{document}

%% file: supp_info_v2/0_laplacian.tex
\section{Laplacian of the positronic density}
\label{sec:laplacian}

\begin{figure}[htbp!]
    \centering
    \includegraphics[width=0.8\linewidth]{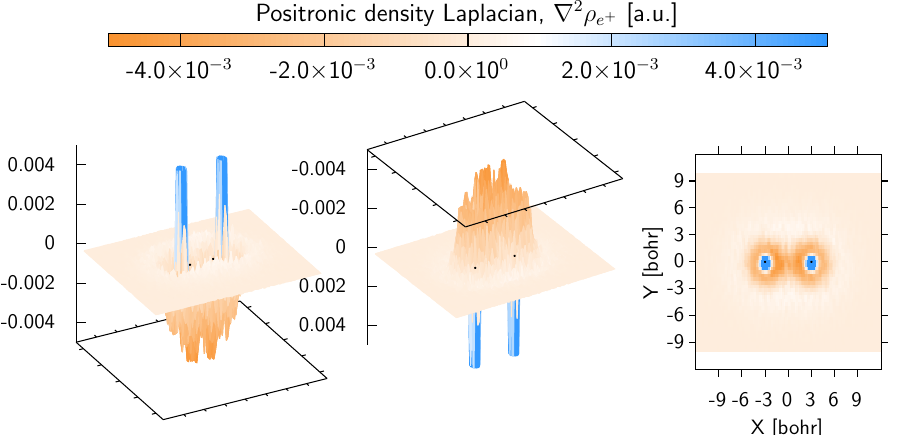}
    \caption{
    Laplacian of the QMC positronic density in (PsH)$_2$. Left: lateral view, Center: inverted lateral, and Right: top view. The QMC 3D density grid was first smoothed via a Gaussian filtering post-processing, then the Laplacian was estimated as three-point finite differences. }
    \label{fig:laplacian}
\end{figure}
\clearpage
\newpage

%% file: supp_info_v2/1_dispersion.tex
\section{Dispersion energy of two non-covalent PsH}
\label{sec:dispersion}

\begin{figure}[htbp!]
    \centering
    \includegraphics[width=0.49\linewidth]{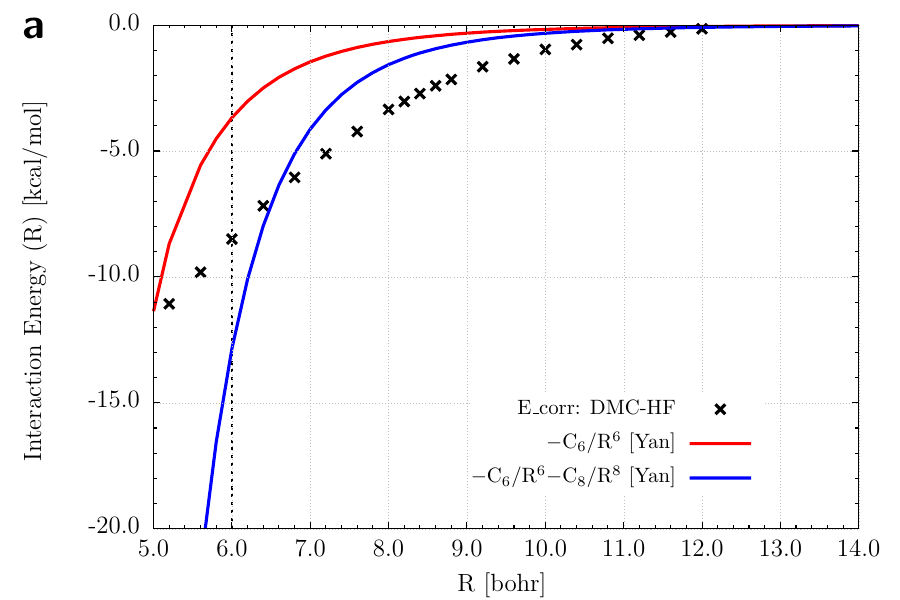}
    \includegraphics[width=0.49\linewidth]{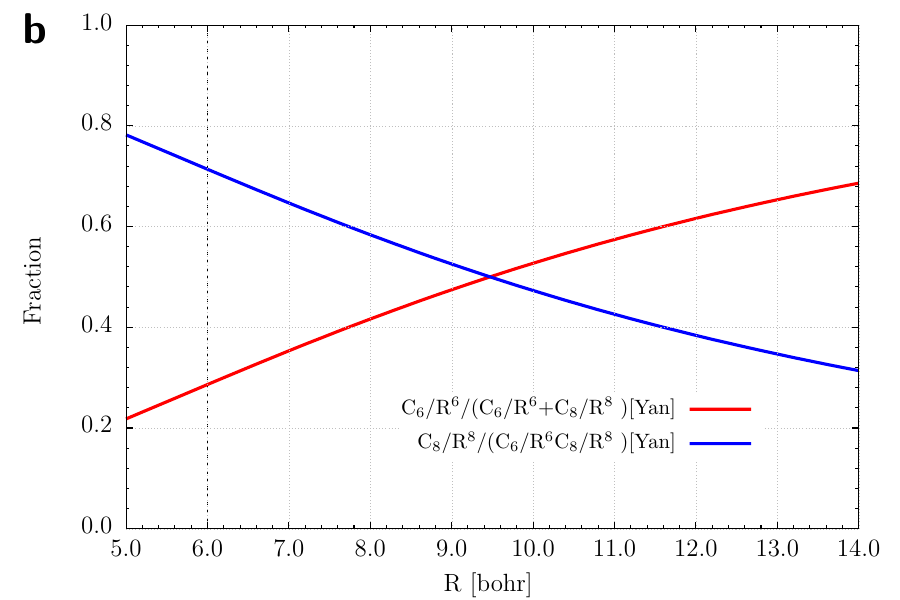}
    \caption{
    (\textbf{a}) Comparison of estimated dispersion energy for PsH-PsH interaction and correlation energy computed from (PsH)$_2$. (\textbf{b}) Fraction of C$_6$ and C$_8$ terms with respect to the total estimated dispersion energy. 
    Dispersion coefficients C$_6$ and C$_8$ taken from Yan et al. \cite{Yan2002}. DMC and HF data from Goli et al\cite{goli_pccp_28_11154_2026}.
Dotted vertical line marks the equilibrium distance of (PsH)$_2$
    Correlation energy defined as the difference between DMC and HF/aug-cc-pVQZ energies $E_{\text{corr}}=E_{\text{DMC}}- E_{\text{HF}}$ 
    }
    \label{SI:fig:dispersion}
\end{figure}

\begin{figure}[htbp!]
    \centering
    \includegraphics[width=0.49\linewidth]{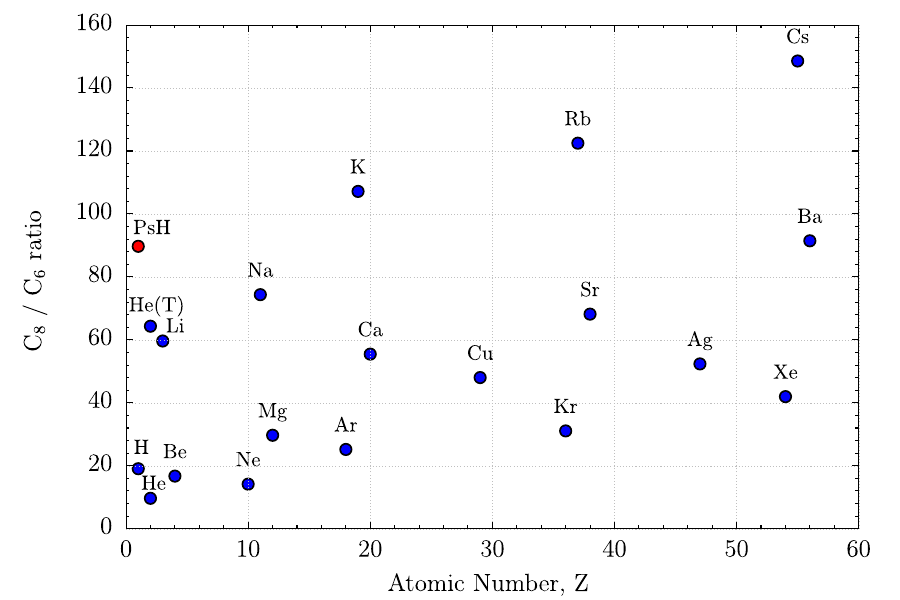}
    \caption{
    Dispersion coefficients $C_8/C_6$ ratio for a set of homonuclear diatomic interactions. 
    Electronic systems data from Jiang et al. \cite{jiang_ADaNDT_101_158_2015}.
    PsH data taken from Yan et al. \cite{Yan2002}.
    }
    \label{SI:fig:c8c6_ratio}
\end{figure}
\clearpage
\newpage

%% file: supp_info_v2/2_quantummontecarlo.tex
\section{Quantum Monte Carlo methods}
\label{sec:qmc}

Quantum Monte Carlo (QMC) methods are a family of stochastic techniques used to integrate the many-body time-independent Schr\"odinger equation over a chosen trial wave function $\Psi_T(\bar{\textbf{r}})$ ~\cite{foulkes_RMP_73_33_2001,Kalos__8_2008, Becca2017}, that can also include explicit correlation between the degrees of freedom of the system.
Here, with the vector $\bar{\textbf{r}}$ of $3N_f$ coordinates, we consider the full set of Cartesian coordinates of the system of electrons and positrons. 

The simplest of these QMC methods, Variational Monte Carlo (VMC)~\cite{foulkes_RMP_73_33_2001,Kalos__8_2008, Becca2017} is based on the stochastic integration of the energy functional of an Hamiltonian $\hat{H}$ over a variational ansatz
$ 
\text{E}\left[\Psi_T\right]=
\frac{\int \Psi^{*}_T(\bar{\textbf{r}}) \hat{\text{H}} \Psi_T(\bar{\textbf{r}})d\bar{\textbf{r}} }{\int \left |\Psi_T(\bar{\textbf{r}}) \right |^2 d\bar{\textbf{r}} } ,
$
which is evaluated by separating the integrand 
$ 
\text{E}\left[\Psi_T\right] = \int E_{\text{loc}}(\bar{\textbf{r}}) \Pi(\bar{\textbf{r}}) d\bar{\textbf{r}},
$ 
into the product of the probability density $\Pi(\bar{\textbf{r}})=\frac{|\Psi_T(\bar{\textbf{r}})|^2}{\int |\Psi_T(\bar{\textbf{r}})|^2 d\bar{\textbf{r}}}$ of finding the system in a configuration $\bar{\textbf{r}}$ and the local energy $\text{E}_{\text{loc}}(\bar{\textbf{r}})=\frac{\hat{\text{H}} \Psi_T(\bar{\textbf{r}})}{\Psi_T(\bar{\textbf{r}})}$, that is the energy of the system in that particular configuration.

By computing the local quantities on a sequence of $\mathcal{N}$ uncorrelated configurations, it is possible to obtain the stochastic estimations of the total quantities, such as the energy $\text{E}_{\text{loc}}(\bar{\textbf{r}})$, through the mean value
$
\text{E}\left[\Psi_T\right]\approx\left\langle \text{E}_{\text{loc}}\right\rangle_\mathcal{N} \pm \sqrt{\frac{\left\langle \text{E}^2_{\text{loc}}\right\rangle_\mathcal{N}-\left\langle \text{E}_{\text{loc}}\right\rangle^2_\mathcal{N}}{\mathcal{N}}},
$
with an error that decreases as the $1/\sqrt{\mathcal{N}}$ depending on the variance $\left\langle \text{E}^2_{\text{loc}}\right\rangle_\mathcal{N}-\left\langle \text{E}_{\text{loc}}\right\rangle^2_\mathcal{N}$.

Within this VMC scheme, we optimize the set of parameters of the wave function, through the Stochastic Reconfiguration procedure described in Refs.~\citenum{sorella_PRB_64_024512_2001, sorella_PRB_71_241103_2005} with the use of correlated sampling technique~\cite{filippi_PRB_61_R16291_2000} in order to better estimate the energy variation between parameter updates. 

Since, in VMC the accuracy in the description of the ground state is still strongly dependent on the trial wave function $\Psi_T(\bar{\textbf{r}};\bar\alpha)$, and so on the dimension of the variational space~\cite{foulkes_RMP_73_33_2001, Kalos__8_2008, Becca2017}.
To obtain a more accurate estimation of the physical observable, better describing the dynamical correlation between the particles in the system, it is common to use the DMC method within the Fixed-Node (FN) approximation~\cite{reynolds_JCP_77_5593_1982} required to overcome the sign problem that arises when dealing with fermionic systems~\cite{kosztin_AJP_64_633_1996, nightingale____1999}.
The fixed node procedure used in this work follows the standard algorithm in ref. \citenum{umrigar_JCP_99_2865_1993} by incorporating several modern features to guarantee better size-consistency properties of the calculations\cite{anderson_JCP_160_104110_2024,dellapia_JCP_163_104110_2025} and stability\cite{barborini_JCP_164_062501_2026}. 

In order to evaluate the properties that do not commute with the Hamiltonian, in this work we used the second order estimator (SOE) corrections to reduce the biasses arising from the mixed averages in FN-DMC\cite{Towler2006,Becca2017}. 

\section{Mixed-Averages and second order estimations in Diffusion Monte Carlo}

In FN-DMC, the time evolution of the product of the trial wave function and the evolved weights generates the mixed distribution $f(\bar{\textbf{r}}, \tau)=\varphi(\bar{\textbf{r}},\tau)\Psi_T(\bar{\textbf{r}})$, where $\varphi(\bar{\textbf{r}},\tau)$ is the best (lower energy) wave function for the ground state with the same nodes as the trial function $\Psi_T(\bar{\textbf{r}})$. 
For the energy and operators that commute with the Hamiltonian operator, this `mixed' estimator corresponds to that obtained on the pure projected state $|\varphi(\bar{\textbf{r}},\tau)|^2$, so that
\begin{equation}
E_{\text{MA}} = \frac{ \langle \Psi_T | \hat{H} | \varphi \rangle }{ \langle \Psi_T | \varphi\rangle  } = 
\frac{ \langle \varphi | \hat{H} | {\varphi} \rangle }{ \langle \varphi | {\varphi} \rangle },
\label{eq:FN_energy}
\end{equation}
For all other operators, that do not commute with the Hamiltonian, the mixed average introduces a bias in their estimations. 
An approximate scheme to reduce this bias, and we show in the work that it is a fundamental correction in all our estimations of the electronic properties, is the second order estimator. 
This approach assumes that the true ground state is close to the trial wave function so that it can be written as\cite{Becca2017} a sum of two terms
\begin{equation}
\varphi=\Psi_T+\delta \Psi,
\label{eq:perturbed_wv}
\end{equation}
where $\delta \Psi$ is normalized and orthogonal to $\Psi_T$. 
In this case the average of a general operator on the projected ground state $\varphi$ becomes
\begin{equation}
\langle\hat{A}\rangle_{\text {pure }}=\left\langle\varphi|\hat{A}| \varphi\right\rangle=\left\langle\Psi_T|\hat{A}| \Psi_T\right\rangle
+2\left\langle\varphi|\hat{A}| \delta \Psi\right\rangle+\langle\delta \Psi|\hat{A}| \delta \Psi\rangle.
\label{eq:pure_average}
\end{equation}
If $\delta \Psi$ is small enough, the second order term $\langle\delta \Psi|\hat{A}| \delta \Psi\rangle$ can be neglected, and after substituting $\left\langle\varphi|\hat{A}| \delta \Psi\right\rangle=$ $\left\langle\Psi_T|\hat{A}| \varphi\right\rangle-\left\langle\Psi_T|\hat{A}| \Psi_T\right\rangle$ we obtain the second order estimator (SOE) as
\begin{equation}
\langle\hat{A}\rangle_{\text{SOE}} \approx 2\langle\hat{A}\rangle_{\text {mix}}-\langle\hat{A}\rangle_{\text {T}}.
\label{eq:estimated_average}
\end{equation}
Within this approximation, it's essential to construct the most accurate possible VMC wave function. 
Which is one of our motivation to develop here the extended dynamical Jastrow factor. 

\clearpage
\newpage

%% file: supp_info_v2/3_hamiltonian.tex
\section{Hamiltonian of the Electron-positrons systems in weak electrostatic fields}\label{sec:hamiltonian}

The metastable positron-molecular states, composed of $N_n$ atomic nuclei, $N_{e}$ electrons (e) and $N_{p}$ positrons (p), at low energies are described within the Born–Oppenheimer approximation through the fermionic Hamiltonian (in atomic units):
\begin{equation}
\hat{\text{H}}_0 = -\sum_{i=1}^{N_{l}} \frac{1}{2}\nabla_{\mathbf{r}_i}^2 +
\sum_{a=1}^{N_n}\sum_{i=1}^{N_l} \frac{Z_a q_i}{|\mathbf{r}_i-\mathbf{R}_a|}  
+\sum_{j>i=1}^{N_l} \frac{q_i q_j}{|\mathbf{r}_i-\mathbf{r}_j|} 
+\sum_{b>a=1}^{N_n} \frac{Z_a Z_b}{|\mathbf{R}_a-\mathbf{R}_b|},
\label{equ:ferm_ham}
\end{equation}
where $N_{l}=N_{e}+N_{p}$ is the total number of leptons, with $N_{e}$ the number of electrons and $N_{p}$ the number of positrons, $N_{n}$ are the number of nuclei in the system, $\{Z_a\}_{a=1}^{N_n}$ are the nuclear charges, $q_i=\pm 1$ are the leptonic charges, and $\bar{\textbf{r}}=\{\textbf{r}_i\}_{i=1}^{N_l}$ and $\bar{\textbf{R}}=\{\textbf{R}_a\}_{a=1}^{N_n}$ are the coordinate vectors of the leptons and nuclei respectively.
In the case of a weak static electric field \cite{Buckingham1967}, the Hamiltonian of the system 
\begin{equation}
\hat{\text{H}} = \hat{\text{H}}_0 - \boldsymbol{\mu} \cdot \textbf{F} 
\label{eq:hamiltonian_field}
\end{equation}
is written as the sum of the unperturbed Hamiltonian in eq. \ref{equ:ferm_ham} plus a term that depends on the electrostatic field vector $\textbf{F}$  and the components of the dipole moment operator $\boldsymbol{\mu}$, which for a system of $N_f$ leptons and $N_n$ nuclei is defined as 
\begin{equation}
    \boldsymbol{\mu} =  \sum^{N_f}_{i=1} q_i \textbf{r}_{i} + \sum^{N_n}_{a=1} Z_a \textbf{R}_{a} .
\label{eq:dipole}
\end{equation}
\clearpage
\newpage

%% file: supp_info_v2/4_wavefunction.tex
\section{Electron-positrons Wave function}
\label{sec:wv}

In the literature, many ans\"{a}tze have been proposed to describe the many-electrons and many-positrons wave function  $\Psi_T(\bar{\textbf{x}}^p, \bar{\textbf{x}}^e; \bar{\textbf{R}})$, where $\bar{\textbf{x}}^e=\{\bar{\textbf{r}}^e,\boldsymbol{\sigma}^e\}$ and $\bar{\textbf{x}}^p=\{\bar{\textbf{r}}^p,\boldsymbol{\sigma}^p\}$ are the set of Cartesian and spin coordinates for the electrons and positrons respectively, and $\bar{\textbf{R}}$ is the vector of the nuclear positions.
Ideally, the most general expression of the wave function should explicitly include all particle correlation effects, including the nuclei, while establishing the correct spatial and spin symmetries for both the electrons and positrons\cite{Bressanini1998, bressanini_JCP_156_154302_2022}. \\
A first approximation to this explicitly correlated wave function was introduced by Bressanini \textit{et al.} in ref. \citenum{Bressanini1998.PRA} and was written as a product of pair correlation functions correctly symmetrized or anti-symmetrized to conform with Pauli's exclusion principle.
In practice, this ansatz attains high accuracy, but its computational cost rises sharply for systems containing large numbers of atoms and positrons, which can limit its applicability to extensive many-body calculations.\\
A way to simplify the total wavefunction is to decouple it into the product 
\begin{equation}\label{eq:elecpos_wvf2}
\Psi(\bar{\textbf{x}}^e,\bar{\textbf{x}}^p; \bar{\textbf{R}}) = 
\psi_e(\bar{\textbf{x}}^e; \bar{\textbf{R}})
\psi_p(\bar{\textbf{x}}^p; \bar{\textbf{x}}^e, \bar{\textbf{R}}) 
e^{\mathcal{J}(\bar{\textbf{x}}^e,\bar{\textbf{x}}^p; \bar{\textbf{R}})} 
\end{equation}
of two fermionic functions, one describing the bound states between the electrons and the nuclei $\psi_e(\bar{\textbf{x}}^e; \bar{\textbf{R}})$ (such as a Slater determinant) and another one describing the coupling between the electrons and the positrons $\psi_p(\textbf{x}^p, \bar{\textbf{x}}^e; \bar{\textbf{R}})$, and a bosonic Jastrow factor.\\
One way to further simplify the wavefunction is to assume that the $\psi_p(\bar{\textbf{x}}^p; \bar{\textbf{x}}^e, \bar{\textbf{R}})$ function is independent of $\bar{\textbf{R}}$ as introduced in ref. \citenum{Bressanini1998} and further generalized for QMC in ref. \citenum{charrymartinez_JCTC_18_2267_2022, barborini_JCP_164_062501_2026}.

Following such an approach, in this work $\psi_e(\bar{\textbf{x}}^e; \bar{\textbf{R}})$ is built as a Slater determinant, whereas the positronic wave function is described by the electron-positron orbitals originally presented in ref. \citenum{charrymartinez_JCTC_18_2267_2022, barborini_JCP_164_062501_2026}, here now is generalized for multi-positronic systems as  
\begin{equation}
\psi_p(\bar{\textbf{x}}^e, \bar{\textbf{x}}^p) = \prod_{i=1}^{N_e} \prod_{j=1}^{N_p} \varphi(\textbf{r}^{ep}_{ij}),
\label{eq:EPO_wv}
\end{equation}
which is based on identical orbitals $\varphi(\textbf{r}^{ep}_{ij})$ (so that the function is symmetric with respect to the exchange of the electronic coordinates) that depend on the electron-positron distance $\textbf{r}^{ep}_{i}$. 
These orbitals $\varphi$ are built as linear combinations 
\begin{equation}
\varphi(\textbf{r}^{ep}) = \sum_{q=1}^{Q} l_q \phi_q\left(\textbf{r}^{ep}\right)
\label{eq:posgem}
\end{equation}
of electron-positron orbitals 
\begin{equation}
\phi\left(\textbf{x}^{ep}\right) = r^{ep} R\left(r^{ep}\right)Y_{l}^{m}(\theta^{ep},\phi^{ep}).
\label{eq:ep_f_orbs}
\end{equation}
Finally, the Jastrow factor consists of the two-body terms between fermions and fermion-nuclei to describe their cusp conditions and pair correlations, together with the full three- and four-body dynamical Jastrow to account for homogeneous and non-homogeneous two-body correlations in the field of the nuclei or positrons as
\begin{equation}
\mathcal{J}(\bar{\textbf{x}}^e,\bar{\textbf{x}}^p; \bar{\textbf{R}}) = 
\mathcal{J}_c(\bar{\textbf{x}}^e,\bar{\textbf{x}}^p;\bar{\textbf{R}}) 
+\mathcal{J}_{3/4}(\bar{\textbf{r}}^e,\bar{\textbf{r}}^p; \bar{\textbf{R}}).
\label{eq:jastrow}
\end{equation}
Since the $\mathcal{J}_c(\bar{\textbf{x}}^e,\bar{\textbf{x}}^p,\bar{\textbf{R}})$ term is extensively discussed in refs. \citenum{charrymartinez_JCTC_18_2267_2022,charry_CS_13_13795_2022, barborini_JCP_164_062501_2026} we will not recall its basic form, on the other hand we will concentrate on the dynamical Jastrow factor $\mathcal{J}_{d}(\bar{\textbf{r}}^e,\bar{\textbf{r}}^p; \bar{\textbf{R}})$ and its extension.

\clearpage
\newpage

%% file: supp_info_v2/5_Jastrow.tex
\section{Three-Four body Jastrow factor for electron-positron correlations}\label{A1:34Jast}

\subsection{Three-/Four- body Jastrow factor}

A first version of the dynamical Jastrow factor for electron-positron systems was introduced in ref. \citenum{charrymartinez_JCTC_18_2267_2022}, inspired by the purely electronic one of Casula \textit{et al.} in ref. \citenum{Casula2004}, as a linear combination of products of atomic orbitals centered on the nuclei.
Although this is not the most common expression of the three-/four-body correlation term\cite{boy+69prs, Drummond2004}, the advantages of this formulation lie in its simplicity and computational efficiency\cite{Casula2004}. 
Yet, this Jastrow explicitly depends only on the distances between the nuclei and the fermions. 
While this is a reasonable correlation factor to describe the attractiveness between nuclei and electrons, it becomes less efficient for the positrons that are actually attracted by the electrons.

In order to further generalize this Jastrow, we thus couple the atomic orbitals with a set of positronic orbitals that explicitly depend on the relative distances between electrons and positrons. 
Therefore, this new dynamical Jastrow factor is built as a combination of three groups of terms
\begin{equation}
\mathcal{J}_{d}(\bar{\textbf{r}}^e,\bar{\textbf{r}}^p; \bar{\textbf{R}}) = 
\mathcal{A}(\bar{\textbf{r}}^e,\bar{\textbf{r}}^p; \bar{\textbf{R}}) 
+\mathcal{G}(\bar{\textbf{r}}^e,\bar{\textbf{r}}^p) 
+\mathcal{M}(\bar{\textbf{r}}^e,\bar{\textbf{r}}^p; \bar{\textbf{R}}).
\label{eq:dyn_Jastrow}
\end{equation} 
which correspond to the atomic Jastrow $\mathcal{A}(\bar{\textbf{r}}^e,\bar{\textbf{r}}^p; \bar{\textbf{R}})$, built purely from the combination of atomic orbitals as previously used in refs. \citenum{charrymartinez_JCTC_18_2267_2022,charry_CS_13_13795_2022}, a set of terms $\mathcal{G}(\bar{\textbf{r}}^e,\bar{\textbf{r}}^p)$ constructed only from the positronic orbitals, and a mixed term that combines both orbital sets, $\mathcal{M}(\bar{\textbf{r}}^e,\bar{\textbf{r}}^p; \bar{\textbf{R}})$.

A schematic representation of all the terms in the Jastrow can be found in Fig. \ref{Fig:Jastrow_factors}.
In the section that follows, we will focus on the justifications that have led to the choice of constructing the three-/four- body Jastrow with all these set of terms. 

\input{supp_info_v2/fig_jastrow}

\subsection{Atomic dynamical Jastrow Factor}\label{ssec:ajf}

The first part of the dynamical Jastrow, already introduced for multi-positronic systems in ref. \citenum{charrymartinez_JCTC_18_2267_2022}, is built on a basis set of $Q$ non-normalized atomic orbitals $\chi_{\nu}(\textbf{r})$, that are centered on different atoms (the atomic index is included in the orbital index $\mu$ or $\nu$ and will always be omitted in the following sections).

For a system comprised of $N_e$ electrons and $N_p$ positrons, this part of the Jastrow factor consists of the sum of three groups of correlation functions,
\begin{equation}
\mathcal{A}(\bar{\textbf{r}}^e,\bar{\textbf{r}}^p; \bar{\textbf{R}}) =
\sum_{j>i}^{N_e} \mathcal{A}^{ee}(\textbf{r}_i^e,\textbf{r}_j^e; \bar{\textbf{R}}) 
+ \sum_{h>t}^{N_p} \mathcal{A}^{pp}(\textbf{r}_h^p,\textbf{r}_t^p; \bar{\textbf{R}})
+ \sum_{i}^{N_e}\sum_{h}^{N_p} \mathcal{A}^{ep}(\textbf{r}_i^e,\textbf{r}_h^p; \bar{\textbf{R}}),
\label{EQ:atm_Jastrow}
\end{equation}
all written as linear combination of products of two atomic orbitals. 
The first set will correlate two electrons 
\begin{equation}
\mathcal{A}^{ee}(\textbf{r}_i^e,\textbf{r}_j^e; \bar{\textbf{R}}) = \sum_{\mu,\nu}^{Q} A^{ee}_{\mu\nu} \chi_{\mu}(\textbf{r}^e_i) \chi_{\nu}(\textbf{r}^e_j) 
\label{EQ:atm_Jastrow_ee}
\end{equation}
the second set will correlate two positrons 
\begin{equation}
\mathcal{A}^{pp}(\textbf{r}_h^p,\textbf{r}_t^p; \bar{\textbf{R}}) = \sum_{\mu,\nu}^{Q} A^{pp}_{\mu\nu} \chi_{\mu}(\textbf{r}^p_h) \chi_{\nu}(\textbf{r}^p_t), 
\label{EQ:atm_Jastrow_pp}
\end{equation}
and the third set will correlate one electron and one positron
\begin{equation}
\mathcal{A}^{ep}(\textbf{r}_i^e,\textbf{r}_h^p; \bar{\textbf{R}}) =  \sum_{\mu,\nu}^{Q} A^{ep}_{\mu\nu} \chi_{\mu}(\textbf{r}^e_i) \chi_{\nu}(\textbf{r}^p_h).
\label{EQ:atm_Jastrow_ep}
\end{equation}
Here the square matrices $\textbf{A}^{ee}$, $\textbf{A}^{pp}$, and $\textbf{A}^{ep}$ that contain the linear coefficients of the expansions, $A^{ee}_{\mu\nu}$, $A^{pp}_{\mu\nu}$, and $A^{ep}_{\mu\nu}$ respectively, describe the coupling between the various atomic orbitals occupied by different fermionic pairs. 
To avoid spin contamination, the $\textbf{A}^{ee}$ and $\textbf{A}^{pp}$ matrices must be symmetric, \textit{ie} $A^{ee}_{\mu\nu}=A^{ee}_{\nu\mu}$ and  $A^{pp}_{\mu\nu}=A^{pp}_{\nu\mu}$, in order for the Jastrow to be invariant with respect to the exchange of two electrons or two positrons, no matter their spin.

Despite the quality of the results presented in refs. \citenum{charrymartinez_JCTC_18_2267_2022,charry_CS_13_13795_2022} these Jastrow terms only contain functions of the electron-nuclei and positron-nuclei distances with no explicit correlation between the two attractive fermionic particles (electrons and positrons).

Yet, the correlation effects arising from explicit electron-positron terms are crucial, especially when describing homogeneous systems of pure electron-positron pairs, or for systems with large numbers of electrons where the positron particle is not in general spherically distributed around an atom\cite{bressanini_JCP_154_224306_2021, bressanini_JCP_155_054306_2021, bressanini_JCP_156_154302_2022, charry_CS_13_13795_2022,charrymartinez_JCTC_18_2267_2022,charry_AC-IE_57_8859_2018}.

For this reason, in the following section we introduce additional components based on electron-positron (or positronic) orbitals\cite{swa+12jcp,charrymartinez_JCTC_18_2267_2022}.


\subsection{Geminal dynamical Jastrow Factor}\label{ssec:gjf}

As previously discussed, the dynamical Jastrow factor constructed as a linear combination of atomic orbitals does not explicitly include the correlation between electron-positron pairs. 
These pair correlation functions are essential, for example, when describing homogeneous systems of pure electrons and positrons, such as the positronium anion (Ps$^-$) or the positronium dimer (Ps$_2$), or in those cases in which, for large molecules, the positrons' wave function is not spherically symmetric around the nuclei.

For this reason, we now introduce additional terms that can be indicated as the Geminal Jastrow factor, which are built through a set of $K$ non-normalized positronic orbitals\cite{charrymartinez_JCTC_18_2267_2022} 
\begin{equation}
\varphi_\eta\left(\mathbf{r}_{ih}^{e p}\right) = r_{ih}^{ep} R\left(r_{ih}^{ep}\right)Y_{l}^{m}(\theta_{ih}^{ep},\phi_{ih}^{ep}),
\label{eq:ep_orbs}
\end{equation}
that depend on the distance vector $\textbf{r}^{ep}_{ih}=\textbf{r}_i^e-\textbf{r}_h^p$ connecting the electron-positron pair, and where $Y_{l}^{m}(\theta_{ih}^{ep},\phi_{ih}^{ep})$ are cubic harmonics, and $R\left(r_{ih}^{ep}\right)$ are non-normalized Gaussian type orbitals. 
For convention, we chose these positronic orbitals to be centered on the positron's position. 

As for the atomic Jastrow, we now describe the three-/four- body correlations between two pairs of electron-positron couples as the sum of three sets of correlation functions
\begin{equation}
\mathcal{G}(\bar{\textbf{r}}^e,\bar{\textbf{r}}^p) = 
\sum^{N_e}_{j>i} \sum^{N_p}_{h} \mathcal{G}^{epe}(\textbf{r}_i^e,\textbf{r}_h^p,\textbf{r}_j^e) 
+\sum^{N_p}_{t>h} \sum^{N_e}_{i}\mathcal{G}^{pep}(\textbf{r}_h^p,\textbf{r}_i^e,\textbf{r}_t^p) 
+\sum^{N_e}_{j>i} \sum^{N_p}_{t>h} \mathcal{G}^{epep}(\textbf{r}_i^e,\textbf{r}_h^p,\textbf{r}_j^e,\textbf{r}_t^p),
\label{EQ:gep_Jastrow}
\end{equation}
each written as the linear combination of products of two positronic orbitals.
Here, the first 
\begin{equation}
\mathcal{G}^{epe}(\textbf{r}_i^e,\textbf{r}_h^p,\textbf{r}_j^e) = \sum_{\eta,\tau}^{K} G^{epe}_{\eta\tau}\varphi_{\eta}(\textbf{r}_{ih}^{ep}) \varphi_{\tau}(\textbf{r}_{jh}^{ep})
\label{EQ:gep_Jastrow_epe}
\end{equation}
and second
\begin{equation}
\mathcal{G}^{pep}(\textbf{r}_h^p,\textbf{r}_i^e,\textbf{r}_t^p)  = \sum_{\eta,\tau}^{K} G^{pep}_{\eta\tau}\varphi_{\eta}(\textbf{r}_{ih}^{ep}) \varphi_{\tau}(\textbf{r}_{it}^{ep})
\label{EQ:gep_Jastrow_pep}
\end{equation}
set of terms represents three fermion correlations respectively between two electrons and one positron (eq. \ref{EQ:gep_Jastrow_epe}) and two positrons and one electron (eq. \ref{EQ:gep_Jastrow_pep}).

Since the Jastrow must be invariant with respect to the exchange of two electronic or positronic coordinates, the square matrices of the coefficients that define the two expansions, $\textbf{G}^{epe}$ and $\textbf{G}^{pep}$, must symmetric, \textit{ie} $G^{epe}_{\eta\tau} = G^{epe}_{\tau\eta}$ and $G^{pep}_{\eta\tau} = G^{pep}_{\tau\eta}$.  

The third set of terms describes the correlation between four different fermions: two electrons and two positrons, again as a product of two geminal orbitals,
\begin{equation}
\mathcal{G}^{epep}(\textbf{r}_i^e,\textbf{r}_h^p,\textbf{r}_j^e,\textbf{r}_t^p) = \sum_{\eta,\tau}^{K} G^{epep}_{\eta\tau} 
\left[ \varphi_{\eta}(\textbf{r}_{ih}^{ep}) \varphi_{\tau}(\textbf{r}_{jt}^{ep}) + \right. 
+\left. \varphi_{\eta}(\textbf{r}_{jh}^{ep}) \varphi_{\tau}(\textbf{r}_{it}^{ep}) \right].
\label{EQ:gep_Jastrow_epep_sym}
\end{equation}
As for the previous two terms, the four-particle correlation function must be symmetric with respect to the exchange of both electrons and positrons, and in order to impose this not only the coupling matrix $\textbf{G}^{epep}$ must be symmetric, \textit{ie} $G^{epep}_{\eta\tau} = G^{epep}_{\tau\eta}$, but also we need to symmetrize the terms with respect to the exchange of both electron pairs and positron pairs, and for this reason we must write them as the sum of the two contributions $\varphi_{\eta}(\textbf{r}_{ih}^{ep}) \varphi_{\tau}(\textbf{r}_{jt}^{ep}) +  \varphi_{\eta}(\textbf{r}_{jh}^{ep}) \varphi_{\tau}(\textbf{r}_{it}^{ep}) $.




\subsection{Mixed atomic-geminal dynamical Jastrow Factor}\label{ssec:mjf}

The atomic and geminal Jastrow factors defined in the previous sections do not exhaust the set of four-body correlations that we can describe with the atomic and positronic basis sets.
While the atomic part neglects the pair correlation between electron-positron pairs, and the Geminal part neglects the influence of the nuclei, we can construct a third term that describes the correlation of two or three fermions in the field of one nucleus, as products of positronic and atomic orbitals.

First, we define the coupling of the $i$th electron with the $h$th positron in the field of the nuclear positions. 
This correlation is constructed as the sum of two combinations, the first of which is the term
\begin{equation}
\mathcal{M}^{nep}(\textbf{r}_i^e,\textbf{r}_h^p; \bar{\textbf{R}})=
 \sum_{\nu}^{Q} \sum_{\eta}^{K} M^{nep}_{\nu\eta} \chi_{\nu}(\textbf{r}_{i}^{e}) \varphi_{\eta}(\textbf{r}_{ih}^{ep}) 
\label{EQ:mxd_Jastrow_nep}
\end{equation}
that correlates the $i$th electron with the nuclear coordinates through the atomic basis set, and with the $h$th positron through the positronic basis set.
The second term inverts the couplings, correlating the positron with the atomic basis and with the electron through the positronic basis set:
\begin{equation}
\mathcal{M}^{epn}(\textbf{r}_i^e,\textbf{r}_h^p; \bar{\textbf{R}})=
 \sum_{\nu}^{Q} \sum_{\eta}^{K} M^{epn}_{\nu\eta} \varphi_{\eta}(\textbf{r}_{ih}^{ep}) \chi_{\nu}(\textbf{r}_{h}^{p}) .
\label{EQ:mxd_Jastrow_epn}
\end{equation}
Following the same procedure we can build terms correlating three fermions in the field of the nuclei. The first set of terms will correlate two electrons and a positron in the field of the nuclei,
\begin{equation}
\mathcal{M}^{enep}(\textbf{r}_i^e,\textbf{r}_j^e,\textbf{r}_h^p; \bar{\textbf{R}})=
\sum_{\nu}^{Q} \sum_{\eta}^{K} M^{enep}_{\nu\eta}
\chi_{\nu}(\textbf{r}_{i}^{e}) \varphi_{\eta}(\textbf{r}_{jh}^{ep}),
\label{EQ:mxd_Jastrow_enep}
\end{equation}
while the second set of terms will correlate two positrons and one electron:
\begin{equation}
\mathcal{M}^{eppn}(\textbf{r}_i^e,\textbf{r}_h^p,\textbf{r}_t^p; \bar{\textbf{R}})=
\sum_{\nu}^{Q} \sum_{\eta}^{K} M^{eppn}_{\nu\eta}\varphi_{\eta}(\textbf{r}_{ih}^{ep}) \chi_{\nu}(\textbf{r}_{t}^{p}).
\label{EQ:mxd_Jastrow_eppn}
\end{equation}
Since the first two sets of terms are two-body correlation functions coupling an electron and a positron no symmetries have to be imposed in the rectangular matrices, $\textbf{M}^{nep}$ and $\textbf{M}^{epn}$.
The same hold for the rectangular matrices $\textbf{M}^{enep}$ and $\textbf{M}^{eppn}$ that appear in the last two three-body terms. 
Yet, here we must guarantee the invariance of the Jastrow factor for the exchange between the two electrons in eq. \ref{EQ:mxd_Jastrow_enep} and the two positrons in eq. \ref{EQ:mxd_Jastrow_eppn}. 
In order to do so, it is sufficient to extend the sums to all pairs of different fermions.
Therefore, using eqs.~\ref{EQ:mxd_Jastrow_epn}, \ref{EQ:mxd_Jastrow_nep}, \ref{EQ:mxd_Jastrow_enep}, and \ref{EQ:mxd_Jastrow_eppn} the total mixed atomic-geminal dynamical factor is expressed as:
\begin{align}
\mathcal{M}(\bar{\textbf{r}}^e,\bar{\textbf{r}}^p; \bar{\textbf{R}}) =& 
\sum^{N_e}_{i} \sum^{N_p}_{h}  \mathcal{M}^{nep}(\textbf{r}_i^e,\textbf{r}_h^p; \bar{\textbf{R}})  
+ \sum^{N_e}_{i} \sum^{N_p}_{h} \mathcal{M}^{epn}(\textbf{r}_i^e,\textbf{r}_h^p; \bar{\textbf{R}})  + \sum^{N_e}_{j \ne i} \sum^{N_p}_{h} \mathcal{M}^{enep}(\textbf{r}_i^e,\textbf{r}_j^e,\textbf{r}_h^p; \bar{\textbf{R}}) \notag \\
&+ \sum^{N_e}_{i} \sum^{N_p}_{t\ne h}  \mathcal{M}^{eppn}(\textbf{r}_i^e,\textbf{r}_h^p,\textbf{r}_t^p; \bar{\textbf{R}}).
\label{EQ:mxd_Jastrow}
\end{align}

In Table \ref{SI:tab:jastrow_basis}, we reported the convergence of the VMC energy for PsH with respect to the size of the atomic and geminal Jastrow basis. Those results help us to establish an adequate compromise between correlation energy gained and computational cost.

\begin{table}[ht!]
\centering
\renewcommand{\arraystretch}{1.0}
\caption{Effect of the size of the atomic Jastrow basis and Geminal electron-positron basis on the VMC energy for PsH (in Hartree units) using the full dynamical Jastrow (atomic, geminal and mixed) see main text. The $\Delta$E column corresponds to the energy difference between each basis set combination and the highest energy obtained.\\}
\label{SI:tab:jastrow_basis}
\begin{tabular}{llcc}
\hline
\multicolumn{1}{c}{\textbf{$\mathcal{A}$}} &
  \multicolumn{1}{c}{\textbf{$\mathcal{G}$}} &
  \multicolumn{1}{c}{\textbf{VMC}} & \multicolumn{1}{c}{\textbf{$\Delta$E}} \\  
\multicolumn{1}{c}{\textbf{Basis}} &
  \multicolumn{1}{c}{\textbf{Basis}} &
  \multicolumn{1}{c}{\textbf{[Ha]}} & \multicolumn{1}{c}{\textbf{[mHa]}} \\ \hline
1s     & 1s     & -0.784278(19) & 0.0  \\
1s     & 3s     & -0.784856(19)  & -0.6 \\
1s     & 3s2p   & -0.784797(18) & -0.5 \\
1s     & 3s2p1d & -0.784794(19) & -0.5 \\
3s     & 1s     & -0.785832(16)  & -1.6 \\
3s     & 3s     & -0.786660(19) & -2.4 \\
3s     & 3s2p   & -0.786674(12) & -2.4 \\
3s     & 3s2p1d & -0.786675(15) & -2.4 \\
3s2p   & 1s     & -0.787104(14) & -2.8 \\
3s2p   & 3s     & -0.787620(15) & -3.3 \\
3s2p   & 3s2p   & -0.788107(12) & -3.8 \\
3s2p   & 3s2p1d & -0.788111(10) & -3.8 \\
3s2p1d & 1s     & -0.787131(14) & -2.9 \\
3s2p1d & 3s     & -0.787630(16) & -3.4 \\
3s2p1d & 3s2p   & -0.788159(9)  & -3.9 \\
3s2p1d & 3s2p1d & -0.788203(13) & -3.9 \\ \hline
\end{tabular}
\end{table}
\clearpage
\newpage

%% file: supp_info_v2/fig_jastrow.tex
\begin{figure}[ht!]
    \centering
    \includegraphics[width=0.9\textwidth]{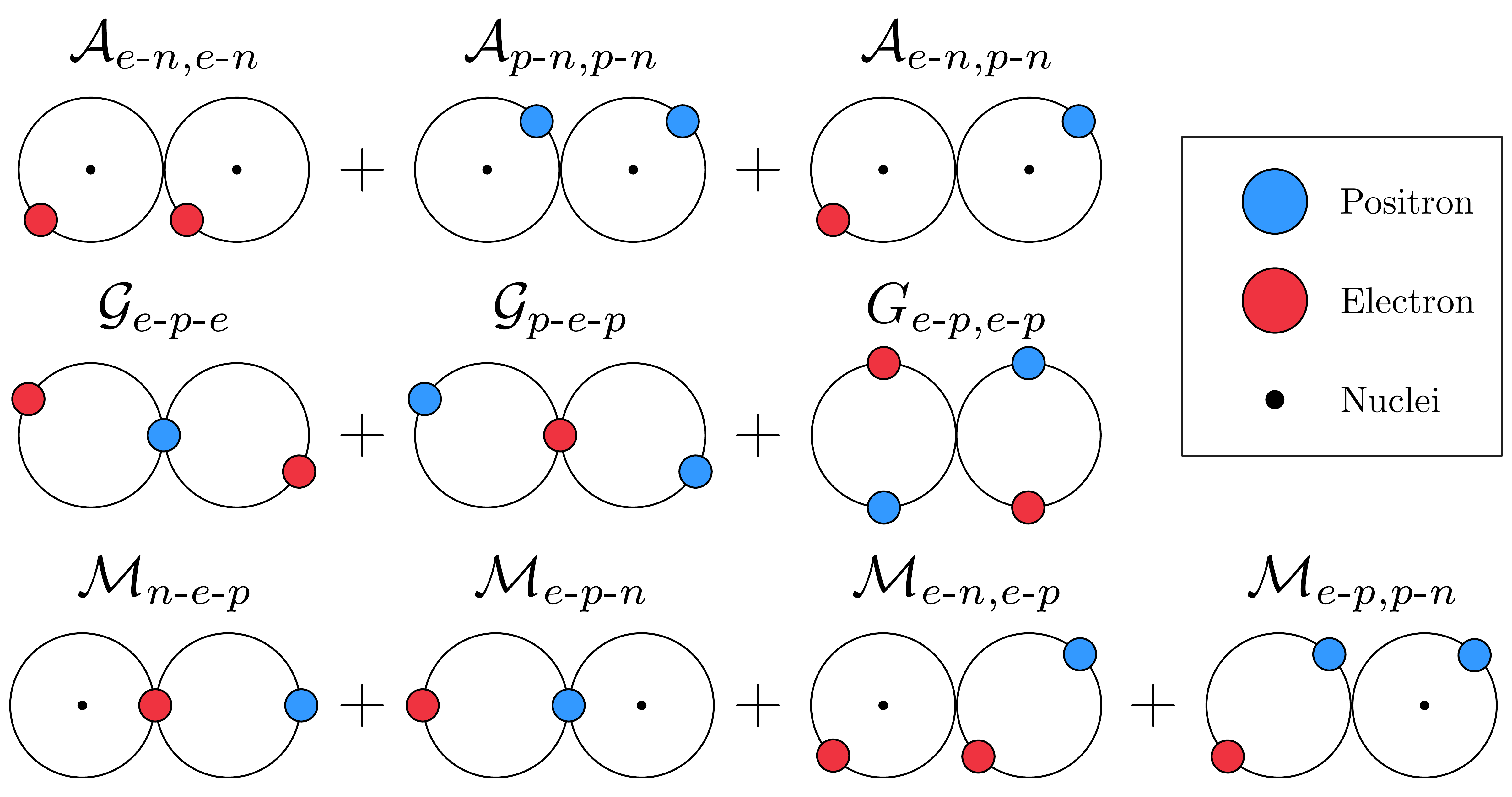}
    \caption{Schematic representation of the dynamical Jastrow factor. $\mathcal{A}$ terms correspond to the Atomic dynamical Jastrow, $\mathcal{G}$ terms for Germinal Electron-Positron dynamical Jastrow, and finally the $\mathcal{M}$ terms for the Mixed Atomic and Geminal dynamical Jastrow. The \textit{hyphen} mark is used to represent the distance between two particles within an orbital and the \textit{comma} symbol to indicate a product of two uncoupled orbitals. }
    \label{Fig:Jastrow_factors}
\end{figure}

%% file: supp_info_v2/6_polarizabilities.tex
\section{External electric field and response properties}
\subsection{External electric field}
In the QMC formalism, the local dipole moment corresponds simply to:
\begin{equation}
\mu_{a,loc}(\bar{\textbf{r}})=\frac{\hat{\mu} \Psi_T(\bar{\lambda};\bar{\textbf{r}})}{\Psi_T(\bar{\lambda};\bar{\textbf{r}})} =  \sum^{N_p}_i q_i r_{i,a}.
\label{eq:dipole_loc}
\end{equation}
Therefore, the local energy corresponding to the Hamiltonian of a system in the presence of a static uniform external electric field can be expressed as 
\begin{align}
E^{F}_{loc}(\bar{\lambda};\bar{\textbf{r}}) &= E^0_{loc}(\bar{\lambda};\bar{\textbf{r}}) - \sum_a \mu_{a,loc}(\bar{\textbf{r}}) F_a 
\label{eq:local_ene_expansion}
\end{align}
Here, it should be noted that the contributions from the external field do not depend on the set of variational parameters $\overline{\lambda}$ of the trial wave function $\Psi_T(\bar{\lambda};\bar{\textbf{r}})$.
\subsection{Polarizabilities}
Considering that the energy of a system under a weak external electric field can be expanded in a power series with respect to the strength of the electric field \cite{Buckingham1967} , in this work we computationally estimated the z-th component, that for the PsH dimer corresponds to the component along the molecular bond, of the dipole polarizabilities by performing a 4th- or 3th-degree polynomial interpolation of the total energies or dipole moment, for a set of different strengths of an external electric field $\vec{F} = \{0,0,F_z\}$ varying from 0.001 to 0.007 Hartree$\cdot$e$^{-1}$$\cdot$bohr$^{-1}$, according to
\begin{equation}
E(F_z) = E_0 -\mu_z^{(0)} F_z -\frac{1}{2!} \alpha_{zz} F_z^2  - \frac{1}{3!} \beta_{zzz} F_z^3  - \frac{1}{4!} \gamma_{zzzz} F_z^4 + \ldots  ,
\label{eq:polar_num_ene}
\end{equation}
where $\gamma$, $\beta$, $\alpha$, $\mu_z^{(0)}$, and $E_0$ are the fitting parameters. Therefore, the second derivative at $\vec{F}=0$ corresponds to the dipole polarizability,
\begin{equation}
\alpha_{ab} = - \left(\frac{\partial^2 E(F_{ab}) }{\partial F^2_{ab}}\right)_{F_{ab} =0}.
\label{eq:dipole_polarizability_e}
\end{equation}
Analogously, the dipole moment can be expanded as 
\begin{align}
\mu_z(\vec{F}) = - \frac{1}{3!} \gamma_{zzzz} F_z^3 - \frac{1}{2!} \beta_{zzz} F_z^2 - \alpha_{zz} F_z - \mu_z^{(0)}
\label{eq:polar_num_dip}
\end{align}
where the first derivative at $\vec{F}=0$ corresponds to the dipole polarizability.
\begin{equation}
\alpha_{ab} = \left(\frac{\partial \mu_b(F_a) }{\partial F_a}\right)_{F_a =0}.
\label{eq:dipole_polarizability_mu}
\end{equation}
For (PsH)$_2$, the only non-spherical symmetric system, the hydrogen nuclei were aligned along the $z$-axis, therefore, the $\alpha_{zz}$ component corresponds to its parallel component $\alpha_{\parallel}$.

\begin{figure}[htbp!]
    \centering
    \includegraphics[width=0.89\linewidth]{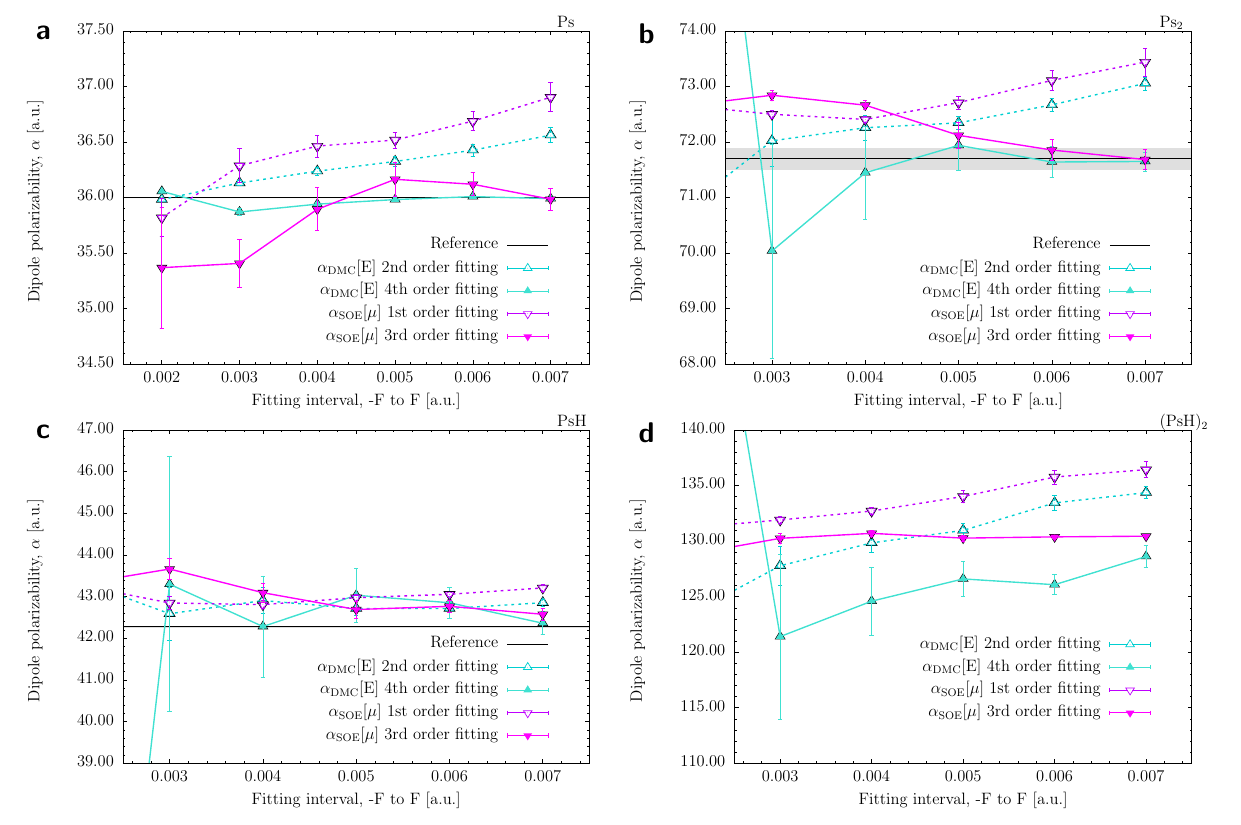}
    \caption{Dipole polarizabilities obtained from fitting of 2nd- and 4th-degree polynomial of the energy, as well as from fitting of 1st- and 3rd-degree polynomial of the total dipole as a function of the strength electric field $F$, for a set of different intervals $-F$ to $F$ spaced every 0.001 a.u. The DMC values computed with a time-step of 0.001 a.u. Ps$2$ reference polarizability from Hylleraas wave function calculations of Ref. \citenum{Yan2002}, and PsH polarizability from PIMC of Ref. \citenum{Tiihonen2018}.  }
    \label{SI:fig:field_convergence}
\end{figure}

Here we performed a systematic analysis on different computation schemes for the dipole polarizability. 
By analyzing the convergence and estimated error with respect to reference values (See figure \ref{SI:fig:field_convergence}), we found that the best performing approach is the 4th-degree polynomial fitting of the energies or 3rd-degree of the dipole, employing the higher interval of electric field strength of 0.007 a.u. 
    
On the other hand, a 2nd-degree polynomial fitting of the energies or 1st- degree of the dipole can also converge to reference values at smaller strengths of the field, but with higher estimated error bars due to the small energy or dipole differences compared to their QMC stochastic error. 

Remarkably, all computed polarizabilites are in agreement within their error bars for all the tested DMC time-steps between 0.001 and 0.015 a.u. (See figure S2). 
This fast convergence suggests that future polarizability studies with QMC could employ the less expensive calculations with higher time steps.

\begin{figure}[htbp!]
    \centering
    \includegraphics[width=0.89\linewidth]{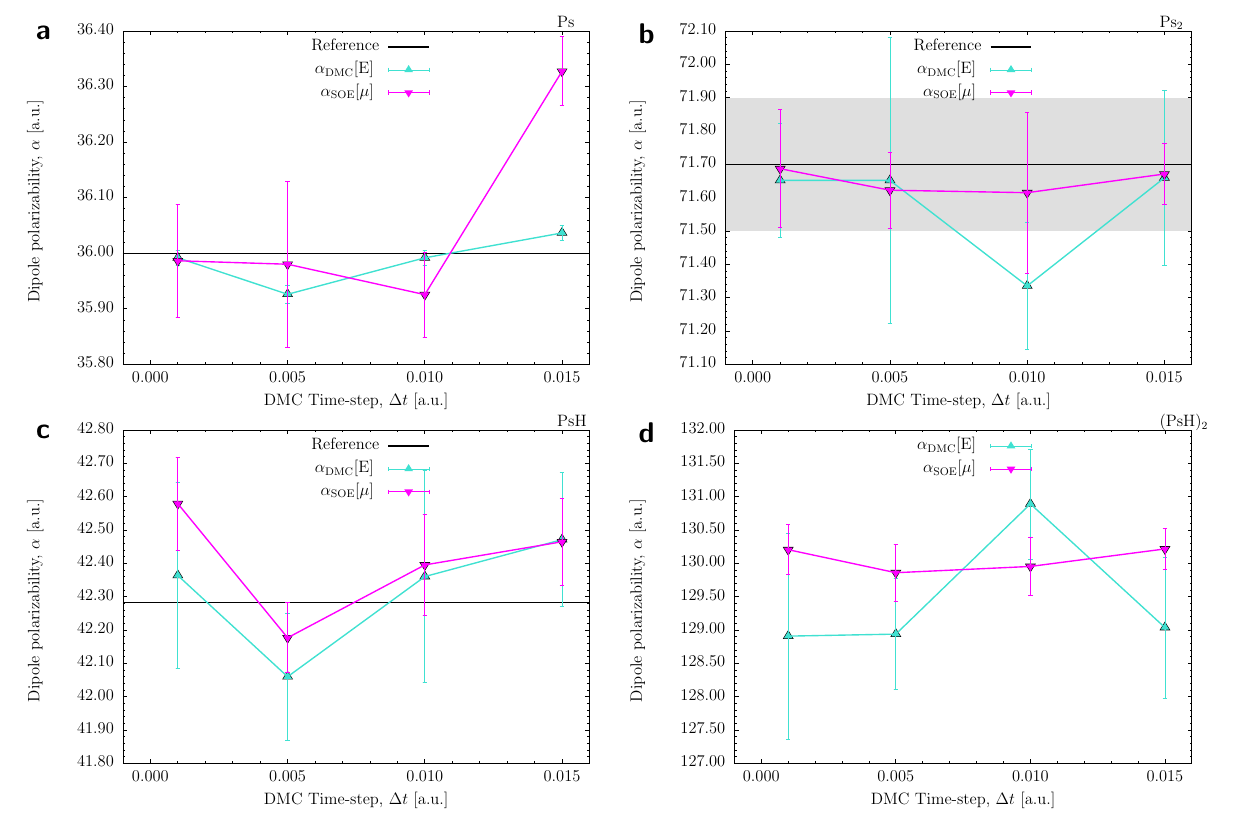}
    \caption{Dipole polarizabilities as a function of the DMC time-step $\Delta t$, obtained from fitting of 4th-degree polynomial of the energy, as well as from fitting of 3rd-degree polynomial of the total dipole as a function of the strength electric field $F$ between -0.007 to 0.007 a.u. Ps$2$ reference polarizability from Hylleraas wave function calculations of Ref. \citenum{Yan2002}, and PsH polarizability from PIMC of Ref. \citenum{Tiihonen2018}.  }
    \label{SI:fig:dt_convergence}
\end{figure}

\begin{figure}[htbp!]
    \centering
    \includegraphics[width=0.89\linewidth]{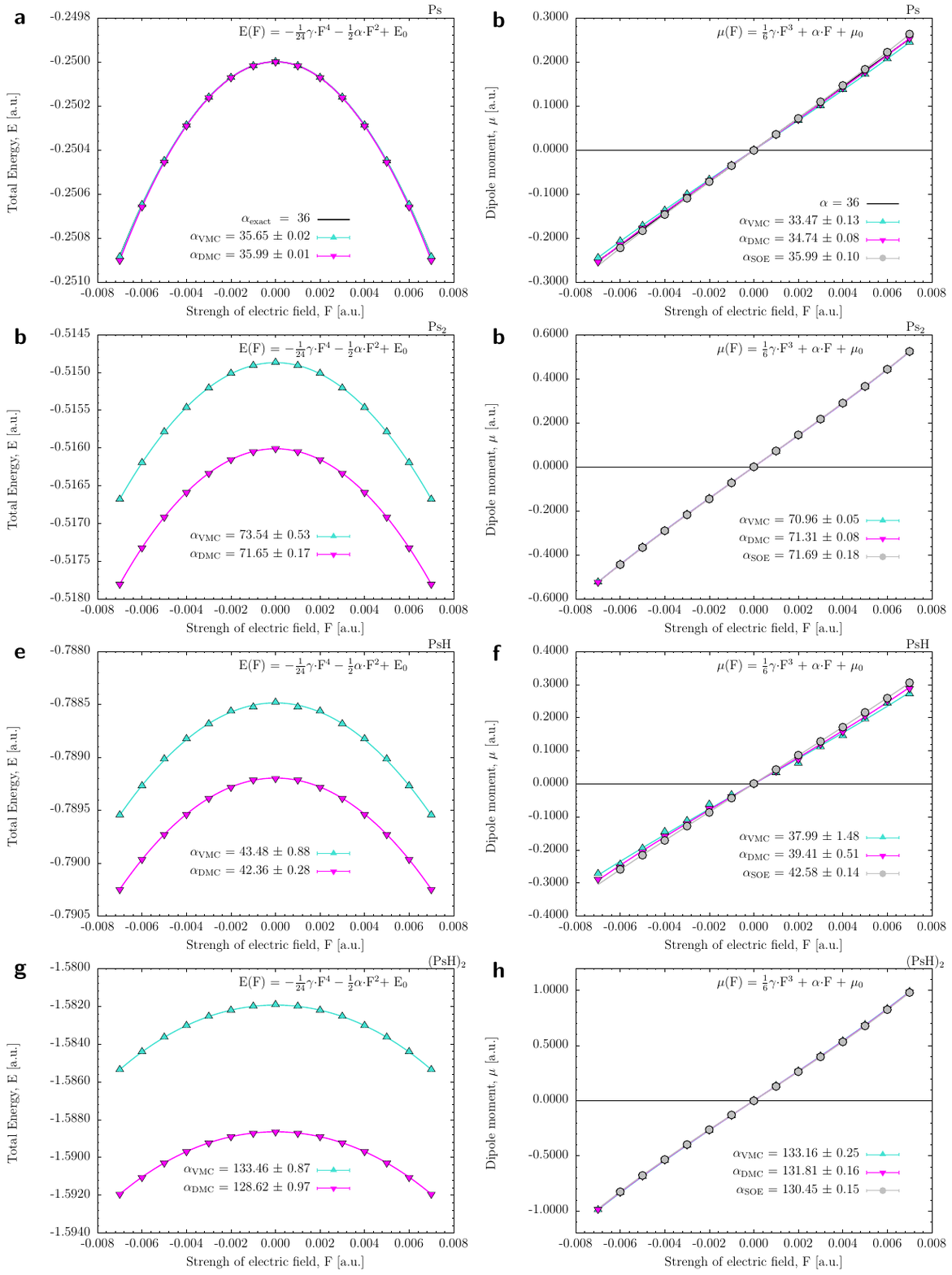}
    \caption{ Total energies (first column) and dipole moment (second column) as a function of the strength of an external electric field for {\textbf{a-b}}: Ps, {\textbf{c-d}}: Ps$_2$, {\textbf{e-f}}: PsH and {\textbf{g-h}}: (PsH)$_2$ (parallel component) computed at VMC (cyan), DMC (magenta), and mixed estimator, MXD (gray). Solid lines indicate the quadratic or linear fit. All error bars are included. }
\label{SI:fig:polarizability}
\end{figure}
\newpage
\pagebreak

\begin{figure}[htpb!]
\centering
\includegraphics[width=0.89\textwidth]{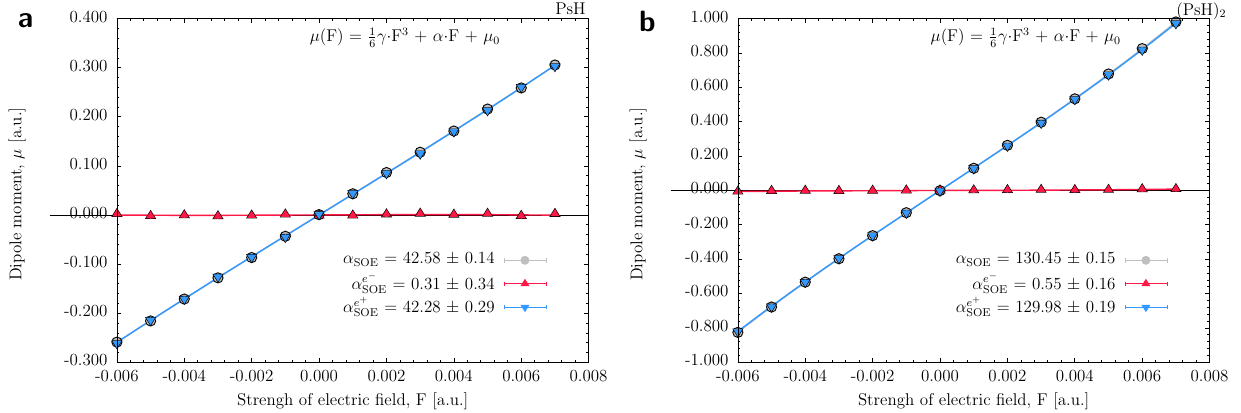}
\caption[Decomposition of dipole polarizability]{Decomposition of the dipole moment (grey) into electronic (red) and positronic (blue) displacement from the nuclei as a function of the strength of an external electric field for \textbf{a}: PsH and \textbf{b}: (PsH)$_2$ (parallel component). }
\label{SI:fig:dipole_decomp}
\end{figure}

\clearpage
\newpage

%% file: supp_info_v2/7_annihilation.tex
\section{Annihilation rate}

For a closed-shell electronic and positronic system, the dominant annihilation path is the two-photon process. Thus, the rate at which an electron and a positron in a singlet state come into direct contact is given by the two-photon annihilation rate \cite{Ferrell1956} 
\begin{equation}
\Gamma_{2}=\pi r_{0}^{2} c\left\langle\delta_{e p}\right\rangle,
\label{eq:annihilation_rate}
\end{equation}
which is proportional to the electron-positron contact density, related to the probability of finding the electron and the positron at a certain point in space 
\begin{equation}
\left\langle\delta_{e p}\right\rangle=\left\langle\Psi\left(\mathbf{r}_{p}, \mathbf{r}_{e}\right)\left|\sum_{i=1}^{N_e}\sum_{j=1}^{N_p} \delta\left(\mathbf{r}_{i}-\mathbf{r}_{j}\right)\right| \Psi\left(\mathbf{r}_{p}, \mathbf{r}_{e}\right)\right\rangle.
\label{eq:ep_probability_density}
\end{equation}
here $c$ is the speed of light in vacuum and $r_0$ is the classical electron radius, which is connected to the Bohr radius $a_0$ by $r_0 = \alpha_{fsc}^2a_0$, where $\alpha_{fsc}$ is the fine structure constant. 

\begin{figure}[htbp!]
    \centering
    \includegraphics[width=0.95\linewidth]{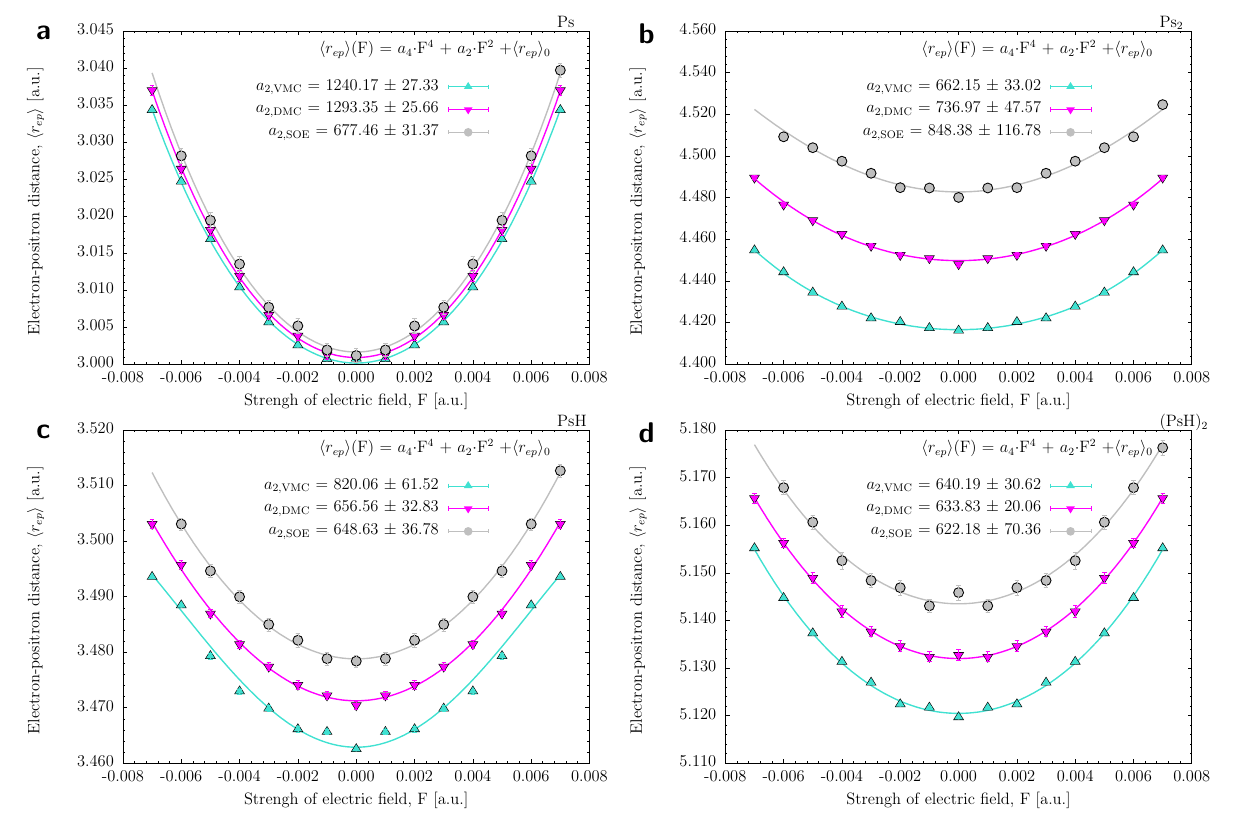}
    \caption{ 
    Electron-positron distances response. Polynomial fitting of the expectation values of the electron-positron distance as a function of the strength of an external electric field, computed with VMC, DMC, and SOE for positronium: Ps (\textbf{a}), positronium dimer: Ps$_2$  (\textbf{b}), positronium hydride: PsH (\textbf{c}), and positronium hydride dimer: (PsH)$_2$ in the parallel component (\textbf{d}). }
    \label{SI:fig:expectation_distances_opposite}
\end{figure}
\begin{figure}[ht!]
    \centering
    \includegraphics[width=0.95\linewidth]{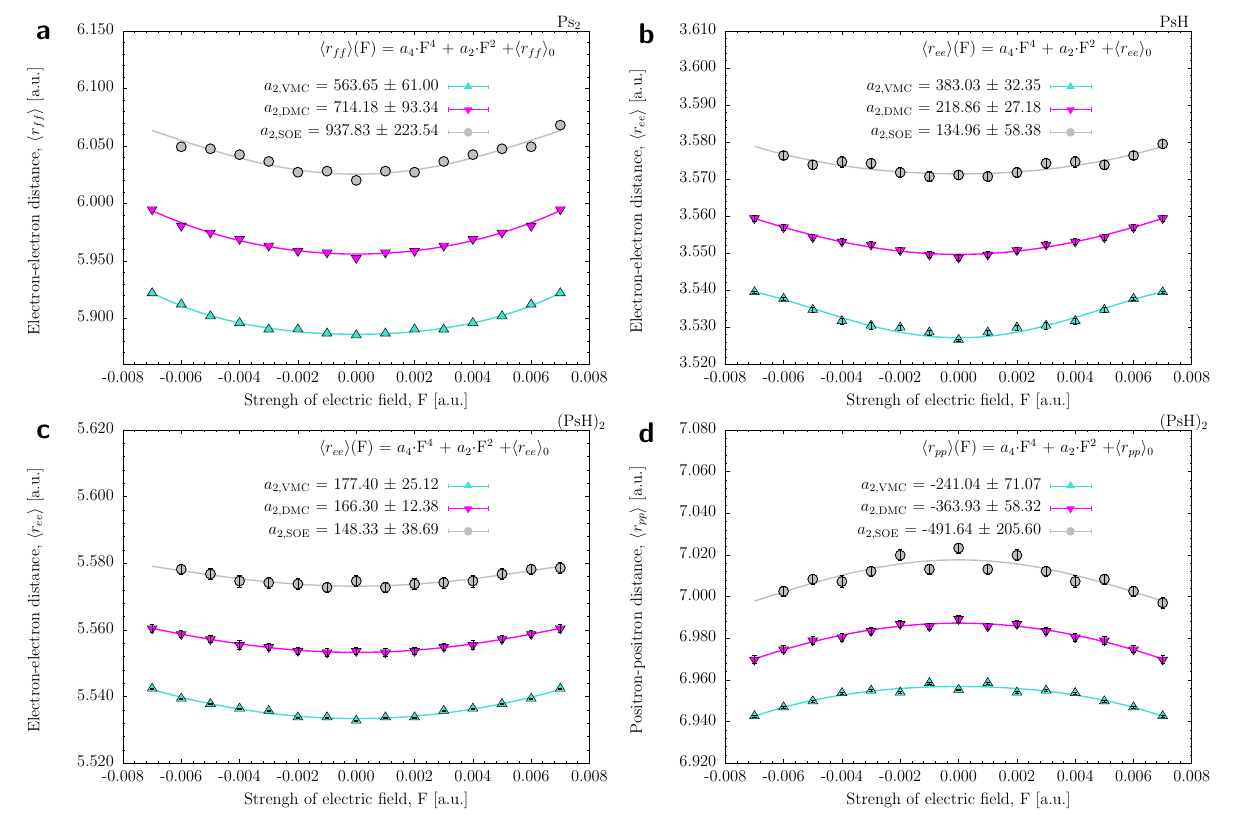}
    \caption{
    Fermion-Fermion distances response. Polynomial fitting of the expectation values of the same type of fermionic particle distance as a function of the strength of an external electric field, computed with VMC, DMC, and SOE for electron-electron (equal to positron-positron) in positronium dimer: Ps$_2$ (\textbf{a}), electron-electron in positronium hydride: PsH (\textbf{b}), electron-electron (\textbf{c}) and positron-positron (\textbf{d})  in positronium hydride dimer: (PsH)$_2$, in the parallel component. }
    \label{SI:fig:expectation_distances_same}
\end{figure}

In QMC, it is extremely expensive to properly sample the coalescence integral in the annihilation rate of equation \ref{eq:ep_probability_density} because of the low number of configurations for which an electron and a positron are located at the same point, or close to it.
For that reason, it is preferred to use the simple extrapolation technique proposed by Jiang \textit{et al.} in Ref. \cite{Jiang1998}, where the coalescence $\delta\left(\mathbf{r}_{i}-\mathbf{r}_{j}\right)$ is replaced by 
\begin{equation}
\delta_{a}\left(x_{i}-x_{j}\right)=\left\{\begin{array}{l}
\frac{3}{4 \pi a^{3}}, \quad \text { for }\left|x_{i}-x_{j}\right| \leqslant a \\
0, \quad \text { otherwise }
\end{array}\right. .
\label{eq:ep_contact_density}
\end{equation}
Therefore, the quantity $\Gamma_2(a)$ should be computed and stored for a set of different distance values $a$ by summing the contribution of each walkers with an electron-positron distance lower than $a$. 
Afterwards, the QMC averages of $\Gamma_2(a)$ are extrapolated to $a=0$ by doing a standard linear fit considering the standard error of the averages.
However, it should be noted that when $a$ is small, the deviation of the counts becomes larger because the counts are few. 

\begin{figure}[htp!]
    \centering
    \includegraphics[width=0.95\linewidth]{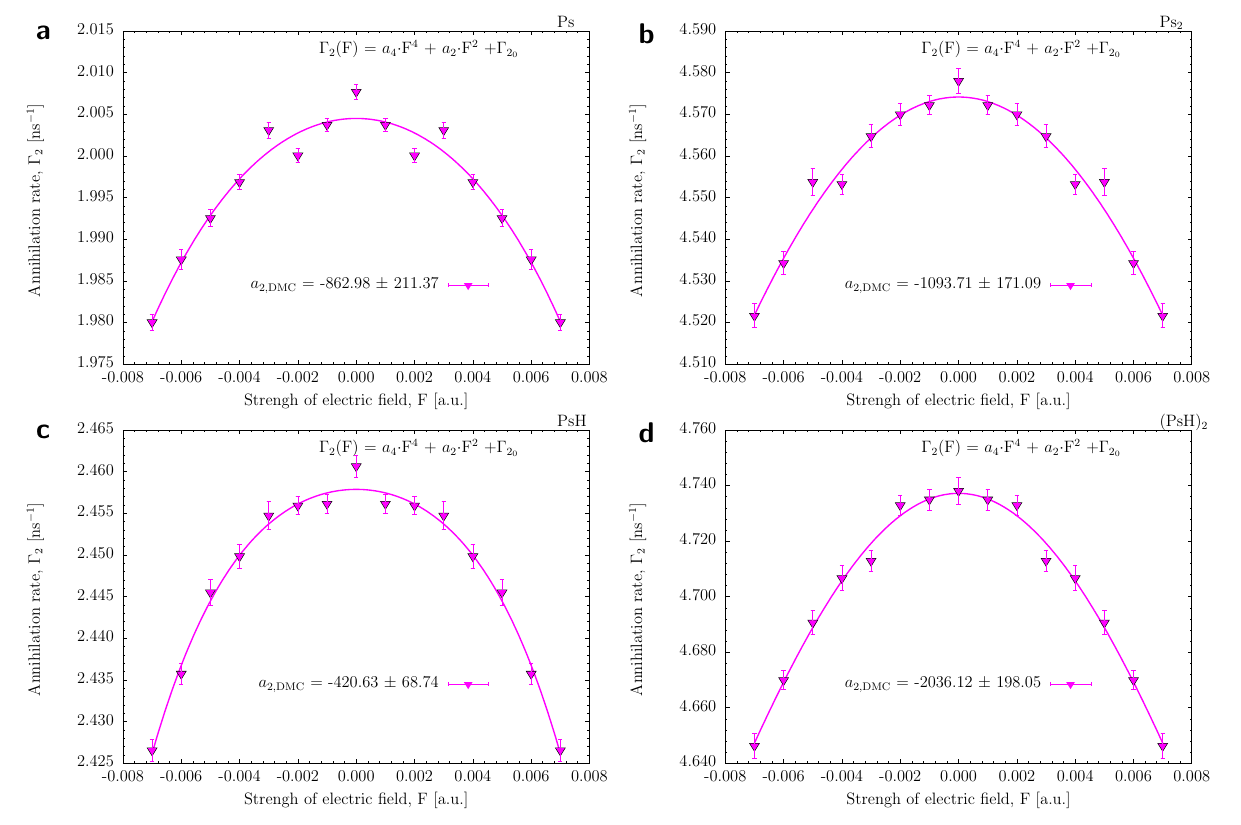}
    \caption{
    Annihilation rates response. Polynomial fitting of the electron-positron pair annihilation rates as a function of the strength of an external electric field, computed with DMC for positronium: Ps (\textbf{a}), positronium dimer: Ps$_2$  (\textbf{b}), positronium hydride: PsH (\textbf{c}), and positronium hydride dimer: (PsH)$_2$ in the parallel component (\textbf{d}). }
    \label{SI:fig:annihilation_rates}
\end{figure}
\pagebreak

\input{supp_info_v2/tab_distances-gamma_full}
\clearpage
\newpage

%% file: supp_info_v2/tab_distances-gamma_full.tex
\begingroup
\begin{table}[htp!]
\centering
\renewcommand{\arraystretch}{1.0}
\caption{Quadratic fitting coefficients, $a_2$ of the polynomial $G(F) = a_0 + a_2F^2 + a_4F^4 $ for expectation values of the distances $\langle r \rangle $ ( in bohr) and annihilation rates (in ns$^{-1}$) as a function of the strength of the external static potential, $F$. \\ }
\label{tab:distances-gamma}
\begin{tabular}{lllll} \hline
                                          & \multicolumn{1}{c}{Ps} & \multicolumn{1}{c}{Ps$_2$} & \multicolumn{1}{c}{PsH} & \multicolumn{1}{c}{(PsH)$_2$} \\ \hline
a$_{2,\text{VMC}} \langle r_{ee} \rangle$ & -         & 564(61)    & 383(32)  & 177(25)    \\
a$_{2,\text{DMC}} \langle r_{ee} \rangle$ & -         & 714(93)    & 219(27)  & 166(12)    \\
a$_{2,\text{SOE}} \langle r_{ee} \rangle$ & -         & 873(211)   & 125(54)  & 156(36)    \\ 
\hline
a$_{2,\text{VMC}} \langle r_{pp} \rangle$ & -         & 564(61)    & -        & -241(71)   \\
a$_{2,\text{DMC}} \langle r_{pp} \rangle$ & -         & 714(93)    & -        & -364(58)   \\
a$_{2,\text{SOE}} \langle r_{pp} \rangle$ & -         & 873(211)   & -        & -479(186)  \\ 
\hline
a$_{2,\text{VMC}} \langle r_{ep} \rangle$ & 620(14)   & 662(33)    & 820(62)  & 640(31)    \\
a$_{2,\text{DMC}} \langle r_{ep} \rangle$ & 647(13)   & 737(48)    & 657(33)  & 634(20)    \\
a$_{2,\text{SOE}} \langle r_{ep} \rangle$ & 677(31)   & 817(110)   & 644(33)  & 632(64)    \\ 
\hline
a$_{2,\text{DMC}} \Gamma_{2}$             & -431(106) & -1094(171) & -421(69) & -2036(198) \\\hline
\end{tabular}
\end{table}
\endgroup

%% file: supp_info_v2/8_general-results.tex
\section{Benchmark on response properties}
\input{supp_info_v2/tab_properties_full}

In Tab. \ref{tab:properties} we benchmark our approach by comparing the Variational Monte Carlo (VMC) and Diffusion Monte Carlo (DMC) for the energies and the Second Order Estimator (SOE), $\langle\hat{A}\rangle_{\text{SOE}} \approx 2\langle\hat{A}\rangle_{\text {DMC}}-\langle\hat{A}\rangle_{\text {VMC}}$, for the electronic and positronic properties, with the results present in the literature. 

First, we can see that from the results reported in Tab. \ref{tab:properties}, the DMC energies are all in agreement with the reference values, within the error bars, while all the expectation values of the interparticle distances, and annihilation rates are in agreement with the references within less than 1$\%$ relative error.
The least accurate estimations are those obtained for the annihilation rates $\Gamma_2$ of Ps$_2$, which have a relative error of 0.9$\%$, due to the larger error in the VMC estimation that greatly influences the SOE (see SM).

Regarding the average interparticle distances $\langle r_{ff} \rangle$(F) and the annihilation rates $\Gamma_2$ in Tab. \ref{tab:properties}, we also report in Tab. \ref{tab:distances-gamma} the non-zero second-degree fitting coefficient $a_2$ used to estimate the zero field quantities 

Such coefficients characterize the variation of the quantities with respect to the external potential, with positive values indicating an increment as a function of the increase of the field's intensity, and negative values indicating the opposite.

In general, the increase of the external field has the effect of increasing the average distances between all the particles, as can be seen for the case of Ps, Ps$_2$ and PsH, and decreasing the annihilation rates $\Gamma_2$, since the increase of $\langle r_{ep} \rangle$(F) induces of reduction of the contact density between the positrons and the electrons.
More specifically, a field of 0.007 a.u. induces a mere change of just 0.005 to 0.05 bohr in the expectation values of the distance, keeping almost unchanged the internal structure of the quantum systems, and a 1$\sim$2$\%$ change in the annihilation rates, which corresponds to a 2$\sim$5 picoseconds increase in the metastable states' lifetimes.

A different behaviour can be seen only in the value of $a_2$ for the $\langle r_{pp} \rangle$(F) distance of the (PsH)$_2$ molecule, which is negative, meaning that the distance between the pairs of positron molecule diminishes as a function of the increase of the field's intensity.
This is due to the particular bond that is formed between the two positrons that mediate the interaction between the two H$^-$ ions.

In Tab. \ref{tab:properties} we report the polarizability of the prototypical conformers.
These are obtained as both the first-order derivative of the dipole moment (using SOE) and then second order derivatives of the energies (see SM for further details in our recommended extrapolation methodology ). 
Importantly, the two estimations coincide within the error bars and with the accurate reference calculations present in the literature\cite{Yan2002,Tiihonen2018}.

For $\text{Ps}_2$, the results confirm the field-free Path-Integral Monte Carlo (PIMC) reference value in ref. \citenum{Tiihonen2018}, which shows that the system has a slightly reduced polarizability compared to the two separate Ps, meaning that its polarizability is less than $2 \alpha_{\text{Ps}}$, despite the large increase in the expected electron-positron average distance. 
For PsH, the estimated value of the polarizability is close to the highly accurate and precise value of Yan \textit{et al.} obtained with extrapolated Hylleraas functions \cite{Yan2002}, corroborating all previous calculations, in contrast to the first reported value of 123 a.u. \cite{LeSech1998}.
\clearpage
\newpage

%% file: supp_info_v2/tab_properties_full.tex
\begingroup
\begin{table}[htp!]
\centering
\renewcommand{\arraystretch}{1.0}
\footnotesize
\caption[Molecular properties of Ps, Ps$_2$, PsH, and (PsH)$_2$]{Dipole polarizabilities (a.u.), expectation values of Interparticle distances (bohr), and two-photon annihilation rates (ns$^{-1}$) for Ps, Ps$_2$, PsH and (PsH)$_2$ systems computed at VMC and DMC levels.\\ }
\label{tab:properties}
\begin{tabular}{lllll} \hline
 & \multicolumn{1}{c}{\textbf{Ps}} & \multicolumn{1}{c}{\textbf{Ps$_2$}} &  \multicolumn{1}{c}{\textbf{PsH}} & \multicolumn{1}{c}{\textbf{(PsH)$_2$}} \\ \hline
E$_\text{VMC}$                        & -0.24999997(1) & -0.51487(1)  & -0.788481(3)        & -1.581927(8) \\
E$_\text{DMC}$                        & -0.2500000(1)  & -0.516007(3) & -0.789199(2)        & -1.58864(1)  \\
E$_\text{ref}$                        & -0.25$^a$      & -0.516004$^b$& -0.789196710(4)$^b$ & -1.5888(1)$^c$   \\ \hline
$\langle r_{ee} \rangle_{\text{VMC}}$ & -              & 5.8852(7)    & 3.5266(3)           & 5.5328(2)    \\
$\langle r_{ee} \rangle_{\text{DMC}}$ & -              & 5.953(1)     & 3.5489(8)           & 5.554(1)     \\
$\langle r_{ee} \rangle_{\text{SOE}}$ & -              & 6.020(2)     & 3.571(1)            & 5.575(2)     \\
$\langle r_{ee} \rangle_{\text{ref}}$ & -              & 6.03321      & 3.574788            & -            \\ \hline
$\langle r_{pp} \rangle_{\text{VMC}}$ & -              & 5.8852(7)    & -                   & 6.9553(4)    \\
$\langle r_{pp} \rangle_{\text{DMC}}$ & -              & 5.953(1)     & -                   & 6.989(2)     \\
$\langle r_{pp} \rangle_{\text{SOE}}$ & -              & 6.020(2)     & -                   & 7.023(2)     \\
$\langle r_{pp} \rangle_{\text{ref}}$ & -              & 6.03321      & -                   & -            \\ \hline
$\langle r_{ep} \rangle_{\text{VMC}}$ & 3.0008(2)      & 4.4161(4)    & 3.4625(2)           & 5.1197(2)    \\
$\langle r_{ep} \rangle_{\text{DMC}}$ & 3.0010(7)      & 4.4481(9)    & 3.4705(7)           & 5.133(1)     \\
$\langle r_{ep} \rangle_{\text{SOE}}$ & 3.0012(10)     & 4.480(1)     & 3.478(1)            & 5.146(2)     \\
$\langle r_{ep} \rangle_{\text{ref}}$ & 3$^a$          & 4.487155$^b$ & 3.480273$^b$        & -            \\ \hline
$\Gamma_{2,\text{VMC}}$               & 2.01(1)        & 4.73(3)      & 2.441(6)            & 4.58(1)      \\
$\Gamma_{2,\text{DMC}}$               & 2.0077(9)      & 4.578(3)     & 2.461(1)            & 4.738(5)     \\
$\Gamma_{2,\text{SOE}}$               & 2.00(1)        & 4.42(3)      & 2.480(7)            & 4.89(1)      \\
$\Gamma_{2,\text{ref}}$               & 2.008$^a$      & 4.465106$^b$ & 2.471406$^b$        & -            \\ \hline
$\alpha_{\text{VMC}}$ [E]             & 35.65(2)       & 73.5(5)      & 43.5(9)             & 133.4(9)     \\
$\alpha_{\text{DMC}}$ [E]             & 35.99(1)       & 71.6(2)      & 42.4(3)             & 129(1)     \\
$\alpha_{\text{VMC}}$ [$\mu$]         & 33.5(1)        & 71.0(1)      & 38(2)               & 133.2(3)       \\
$\alpha_{\text{DMC}}$ [$\mu$]         & 34.74(8)       & 71.3(1)      & 39.4(5)             & 131.8(2)       \\
$\alpha_{\text{SOE}}$ [$\mu$]         & 36.0(1)        & 71.7(2)      & 42.6(1)             & 130.5(2)       \\
$\alpha_{\text{ref}}$                 & 36$^{d}$     & 71.7(2)$^d$  & 42.2836(5)$^e$      & -            \\ \hline
$\alpha^{e}_{\text{SOE}}$ [$\mu^{e}$] & 18.0(1)    & 35.8(5)      & 0.3(3)              & 0.5(2)       \\
$\alpha^{p}_{\text{SOE}}$ [$\mu^{p}$] & 18.0(1)    & 35.8(5)      & 42.3(4)             & 130.0(2)       \\  \hline
\end{tabular} \\
$^a$ Ps analytical values from ref. \citenum{mitroy_NDiACaP__199_2002}
$^b$ Explicitly correlated gaussians (ECG) from ref. \citenum{Bubin2006}
$^c$ DMC \cite{bressanini_JCP_155_054306_2021}
$^d$ Path Integral Monte Carlo (PIMC) from ref. \citenum{Tiihonen2018}
$^e$ Hylleraas wave function from ref. \citenum{Yan2002} 
\end{table}
\endgroup

%% file: supp_info_v2/9_PES.tex
\section{Potential energy curves and polarizability curves}
\begin{figure}[htp!]
    \centering
    \includegraphics[width=0.80\linewidth]{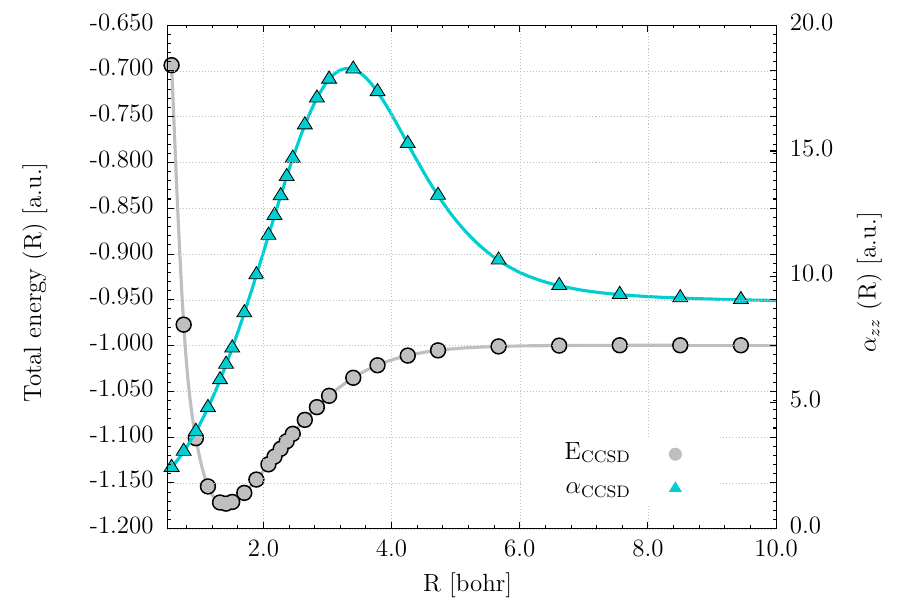}
    \caption{Total energy and dipole polarizability for H$_2$ computed at CCSD/aug-cc-pVTZ level}
    \label{fig:pes_h2}
\end{figure}

\begin{figure}[htp!]
    \centering
    \includegraphics[width=0.80\linewidth]{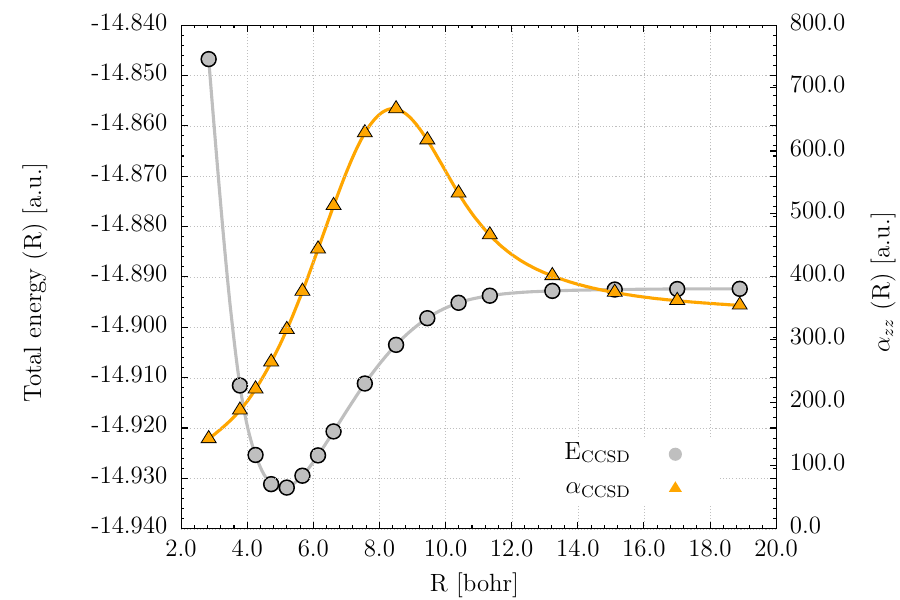}
    \caption{Total energy and dipole polarizability for Li$_2$ computed at CCSD/aug-cc-pVTZ level}
    \label{fig:pes_li2}
\end{figure}

\begin{figure}[htp!]
    \centering
    \includegraphics[width=0.80\linewidth]{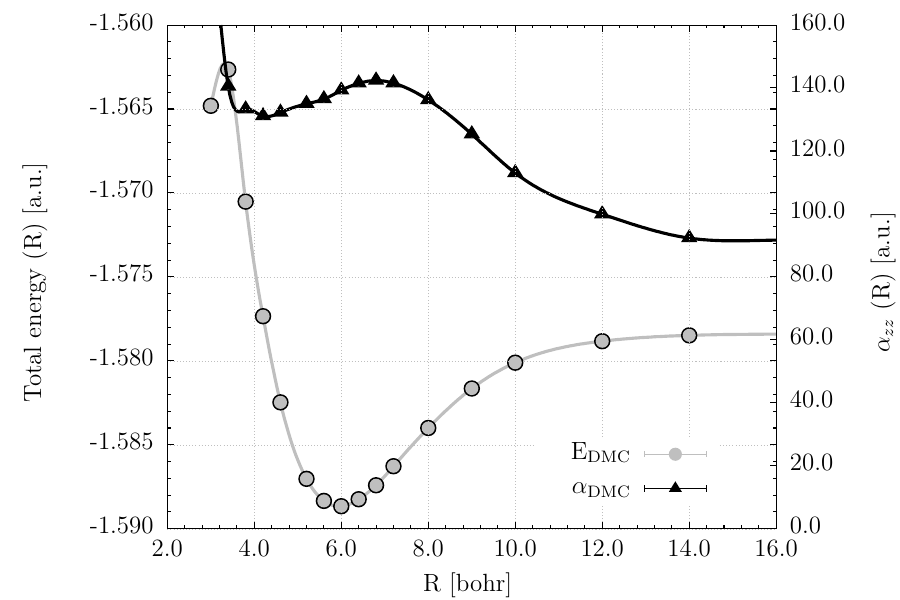}
    \caption{Total energy and dipole polarizability for (PsH)$_2$ computed at DMC level with a 0.001 a.u. time step}
    \label{fig:pes_psh2}
\end{figure}

\begin{figure}[htp!]
    \centering
    \includegraphics[width=0.80\linewidth]{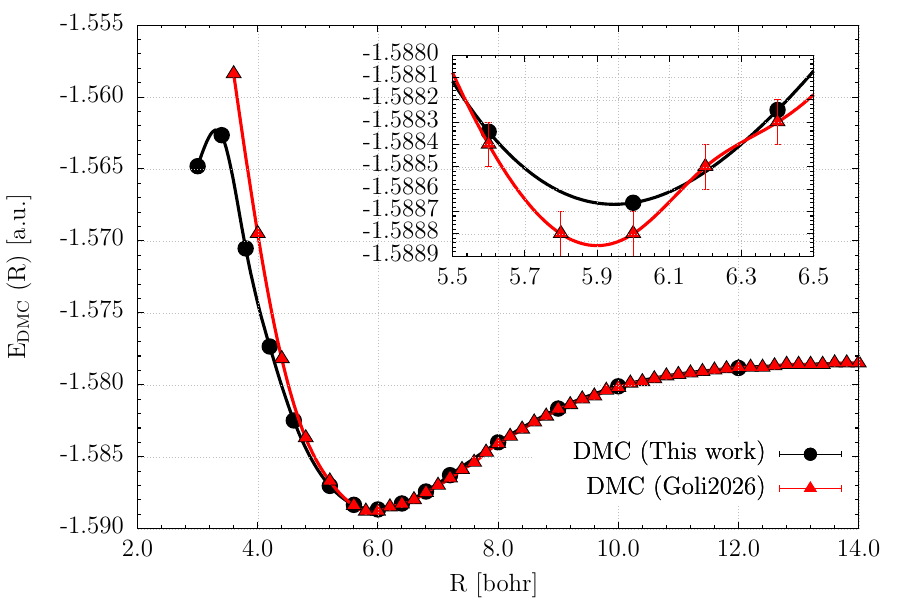}
    \caption{Comparison of the DMC energies for (PsH)$_2$. Goli2026 data was taken from reference \cite{goli_pccp_28_11154_2026}, computed at DMC level with a 0.003 a.u. time step }
    \label{fig:pes_psh2_comp}
\end{figure}

\begin{figure}[htp!]
    \centering
    \includegraphics[width=0.80\linewidth]{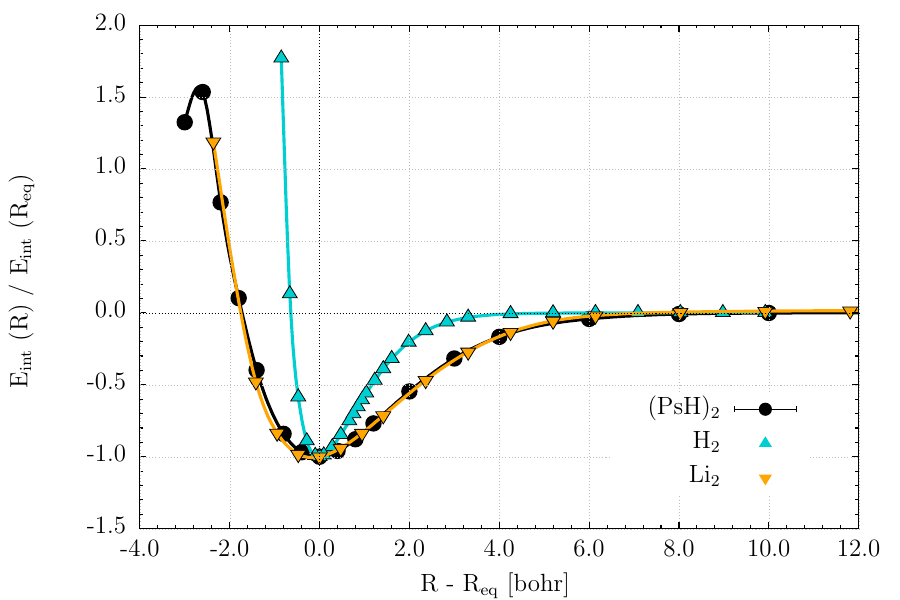}
\caption[PsH2 PES]{Relative interaction energies as a function of the internuclear separation shifted to the equilibrium for (PsH)$_2$ (black circles), H$_2$ (turquoise up-triangles), and Li$_2$ (orange down-triangles)}
\label{fig:pes_relative}
\end{figure}

\begin{figure}[htp!]
    \centering
    \includegraphics[width=0.8\linewidth]{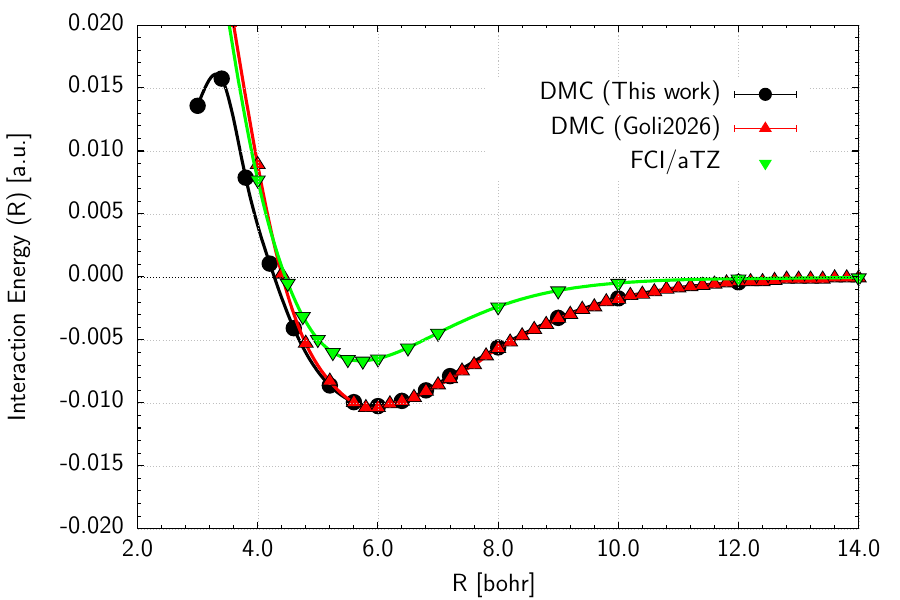}
    \caption{
    Comparison of the interaction energies of (PsH)$_2$ at QMC and Full Configuration Interaction (FCI/aug-cc-pVTZ) with a reduced active space selected from the natural orbitals of a preceding CISD calculation. 
    }
    \label{fig:psh2_comp_fci}
\end{figure}

\clearpage
\newpage